\documentclass[aps,prl,twocolumn,superscriptaddress,groupedaddress]{revtex4}  
\usepackage{graphicx}  
\usepackage{dcolumn}   
\usepackage{bm}        
\usepackage{amssymb}   
\usepackage{multirow}
\usepackage{epsfig}
\usepackage{amsmath}
\usepackage{float}

\begin{document}

\lefthyphenmin=2
\righthyphenmin=2

\widetext

\title{A novel method for lepton energy calibration at Hadron Collider Experiments}
\affiliation{University of Iowa, Iowa City IA, United States of America}
\affiliation{Department of Modern Physics, University of Science and Technology of China, Anhui, China}
\author{Siqi Yang}\affiliation{University of Iowa, Iowa City IA, United States of America}
\author{Usha Mallik}\affiliation{University of Iowa, Iowa City IA, United States of America}
\author{Liang Han}\affiliation{Department of Modern Physics, University of Science and Technology of China, Anhui, China}
\author{Weitao Wang}\affiliation{Department of Modern Physics, University of Science and Technology of China, Anhui, China}
\author{Jun Gao}\affiliation{Department of Modern Physics, University of Science and Technology of China, Anhui, China}
\author{Minghui Liu}\affiliation{Department of Modern Physics, University of Science and Technology of China, Anhui, China}

\date{March. 05, 2018}
\begin{abstract}
  This report is to provide a novel method for the lepton energy calibration at Hadron Collider Experiments. 
  The method improves the classic lepton energy calibration procedure widely used at hadron 
  collider experiments. The classic method parameterizes the potential 
  bias in the lepton energy calibration, and determines the value of the parameter by the invariant mass 
  of $Z/\gamma^*\rightarrow \ell^+\ell^-$ events. The precision of the calibration is dominated by the number of 
  parameters or terms considered in the parameterization, for example, a polynomial extension. 
  With one physics constraint of the reconstructed $Z$ boson mass, the classic procedure can use and 
  determine one parameter. 

  The novel method improves the precision of lepton calibration by introducing more terms in the parameterization. 
  To precisely determine the values of multiple parameters, the method first acquires 
  more constraints by separating the $Z/\gamma^*\rightarrow \ell^+\ell^-$ samples according to the decay kinematics, and then 
  reduces the correlation between multiple parameters. Since the new method is still using the 
  reconstructed $Z$ boson masses as the only constraints, it is much faster and easier than 
  detailed study of detector simulations. 
\end{abstract}
\maketitle

\section{I. Introduction}
\subsection{I-A. Lepton energy calibration}\label{sec:classic}

  Measurement and reconstruction of electron and muon energy at hadron collider experiments are 
  essential in many physics analyses.   
  For experiments like D0 and CDF from the Fermilab Tevatron, and 
  ATLAS and CMS from the CERN Large Hadron Collider (LHC), the determination of electron and 
  muon energy scale is required to have a high precision for a wide range of energy from a few GeV up to 
  $\mathcal{O}(1000)$ GeV.
  After various calibrations, the corrected lepton energy scale is required to be consistent with its 
  true value:
  
\begin{eqnarray}\label{eq:requirement}
  \mathcal{E}(E_\text{obs}) &\rightarrow&  \mathcal{E}(E_\text{corr}) = \mathcal{E}(E_\text{true})
\end{eqnarray}
  where $E_\text{obs}$ and $E_\text{corr}$ are the observed lepton energy and calibrated lepton energy.
  $E_\text{true}$ is the corresponding true energy. $\mathcal{E}$ means average or mathematical expectation on an 
  ensemble of events, representing 
  the energy scale.
  To satisfy this requirement, a widely used method has been developed 
  at hadron collider experiments, to determine and calibrate electron and muon energy scale 
  using the $Z/\gamma^*\rightarrow \ell^+\ell^-$ (dilepton) events. 
  In this method, the lepton energy scale is determined by observing the dilepton invariant mass instead of 
  directly looking into the energy spectrum. 
  The reconstructed dilepton mass 
  can be expressed as:
  
\begin{eqnarray}\label{eq:originalMass}
  M^2 = 2E_1 E_2 (1-\cos\theta_{12}), 
\end{eqnarray}
  where $E_1$ and $E_2$ are the energy of the two leptons, and $\theta_{12}$ is the spatial opening angle 
  between them. 
  The invariant mass observed in the dilepton  
  events, which is determined by the line shape of the $Z$ boson invariant mass spectra, has a very sharp peak, thus 
  sensitive to the lepton energy scale.
  A scaling factor $k$ is applied to the lepton energy as a correction:

\begin{eqnarray}\label{eq:singleEcorr}
  E_\text{corr} &=& k\cdot E_\text{obs}.
\end{eqnarray}
  According to Eq.~\ref{eq:originalMass}, we have

\begin{eqnarray}\label{eq:masscalib}
  M^2_\text{corr} &= & k^2 M^2_\text{obs}, 
\end{eqnarray}
  Therefore, the energy calibration in Eq.~\ref{eq:requirement} is 
  equivalent to a mass calibration:
  
\begin{eqnarray}\label{eq:massrequirement}
  \mathcal{E}(M_\text{obs}) \rightarrow \mathcal{E}(M_\text{corr}) &=& \mathcal{E}(M_\text{true}) \nonumber \\
     &=& k \cdot \mathcal{E}(M_\text{obs}), 
\end{eqnarray}
  which means the correction $k$ on lepton energy scale can be determined by requiring the best agreement 
  between the corrected 
  mass mean $\mathcal{E}(M_\text{corr})$, 
  and the ``true'' value $\mathcal{E}(M_\text{true})$. 
  Most physics measurements are unbiased as long as energy scales in data and Monte Carlo (MC) simulations 
  are consistent. Thus $\mathcal{E}(M_\text{obs})$ and $\mathcal{E}(M_\text{true})$ are the mass means observed in 
  MC and data. 
  Sometimes, physics measurements 
  requires not only a consistent energy scale, but also an absolute energy calibration. In this case, $\mathcal{E}(M_\text{true})$ 
  indeed represents the ``true'' value which can be acquired from generator level information.

  The energy scale is determined by observing dilepton mass instead of  lepton energy because of two main reasons. 
  First, the overall uncertainty on the energy scale determination is dominated by the uncertainty of observing the mean value. 
  The uncertainty of observing mean value from any gaussian-type sample is expressed as:
  
\begin{eqnarray}\label{eq:statunc}
  \frac{\delta \text{mean}}{\text{mean}} = \frac{\Delta}{\text{mean}\cdot \sqrt{N}}
\end{eqnarray}
  where $\Delta$ is the standard deviation of the sample, and $N$ is the total number of events 
  in this sample. 
  As we know, the width of the $Z$ boson mass spectra is only a few GeV (including physics width and 
  resolution of detector), while the width of the lepton energy distribution can be very large at hadron 
  colliders. Therefore, the statistical uncertainty on $k$ factor is much smaller from the mass mean observation.

  Second, the main contribution of the observed lepton energy at hadron colliders comes from the $Z$ boson boost and 
  the finite $p_T$ of the $Z$ recoiling against some hadronic system. 
  rather than the $Z$ boson mass. 
  The $Z$ boson boost is determined by the difference of energy between initial 
  quarks and anti-quarks of $q\bar{q}\rightarrow Z/\gamma^*$ annihilation, where quarks come from hadrons. 
  It means parton distribution 
  functions (PDFs) used in MC generation has impact. QCD calculation of initial state 
  radiation also affects the lepton energy in MC. 
  As a result, large effects from PDFs and QCD calculation will be absorbed in the difference between 
  $\mathcal{E}(E_\text{obs})$ and $\mathcal{E}(E_\text{true})$, and further propagate to the 
  energy scale determination.
  By observing invariant mass instead, the calibration is independent of PDFs and QCD calculation.
  It not only reduces the systematic uncertainty, but also is a necessary condition of absolute energy calibration. 
  
  In general, this method does not depend upon knowledge on understanding the source of the bias 
  in energy measurement, such as multiple hadron interactions (pileup effects), energy loss and imperfect 
  reconstruction algorithm. 
  All these effects are absorbed in the $k$ factor as a simple parameterization. 
  It is the reason that calibration using observed dilepton 
  events is always applied as the final step after all detector and simulation level calibrations at hadron collider experiments. 

  Since the data accumulates very fast at hadron colliders with high luminosity, the $Z/\gamma^*\rightarrow \ell^+\ell^-$ sample 
  can be very large. According to Eq.~\ref{eq:statunc}, 
  a precision of lepton energy calibration of $0.01\%$ should be easily achieved using 10 M dilepton  
  events, corresponding to a data sample collected by the ATLAS or the CMS detector within one year at 13 TeV~\cite{ATLAS-Z-xsection}.
  
\subsection{I-B. Limited precision}

  In spite of an expected high precision, the calibration procedure described in section I-A has a much larger uncertainty 
  which will not be reduced as data accumulating. 
  As discussed, the calibration using dilepton events 
  is a parameterization of any potential bias in the lepton energy measurement. A perfect parameterization 
  can be written in the form of a polynomial expansion:
  
\begin{eqnarray}\label{eq:fullEcorr}
  E_\text{truth} = b + k\cdot E_\text{obs} + \gamma\cdot E^2_\text{obs}+ \cdot\cdot\cdot ,
\end{eqnarray}
  Higher order terms in Eq.~\ref{eq:fullEcorr} can be very small with modern detector design and construction, but 
  offset term $b$ could be very large due to noise from pileup effects, resulting in a linear 
  relationship between observed energy and its true value:
  
\begin{eqnarray}\label{eq:linearErelation}
  E_\text{truth} = b + k \cdot E_\text{obs}.
\end{eqnarray}
  However, the calibration in Eq.~\ref{eq:singleEcorr} is using a single scaling parameterization, here denoted as $k'$
  
\begin{eqnarray}\label{eq:wrongErelation}
  E_\text{obs} \rightarrow E_\text{corr} &=& k' \cdot E_\text{obs}, 
\end{eqnarray}
  and still requires $\mathcal{E}(M_\text{true}) = \mathcal{E}(M_\text{corr})$. Thus we 
  have:
  
\begin{eqnarray}\label{eq:singlefactor_avg}
  k' \cdot \mathcal{E}(E_\text{obs}) &\approx& b + k \cdot \mathcal{E}(E_\text{obs})  \nonumber \\
  k' - k &\approx& \frac{b}{\mathcal{E}(E_\text{obs})}, 
\end{eqnarray}  
  Note that when $k'$ is determined by observing mass, and $b$ is not negligible, 
  Eq.~\ref{eq:singlefactor_avg} can only hold as an approximation, which will 
  be discussed later in section II-C. 
  An energy-dependent bias then appears as:
  
\begin{eqnarray}\label{eq:runningunc}
 \frac{E_\text{corr} - E_\text{truth}}{E_\text{obs}} &=& \frac{k'\cdot E_\text{obs} - k \cdot E_\text{obs} - b}{E_\text{obs}} \nonumber \\
   &=& (k' - k) \cdot \left[ 1 - \frac{\mathcal{E}(E_\text{obs})}{E_\text{obs}} \right] \nonumber \\
   &=& \frac{b}{\mathcal{E}(E_\text{obs})}\left[ 1- \frac{\mathcal{E}(E_\text{obs})}{E_\text{obs}} \right].
\end{eqnarray}
  The dependence is determine by $b/\mathcal{E}(E_\text{obs})$. For the $pp\rightarrow Z/\gamma^*\rightarrow \ell^+\ell^-$ events 
  at $\sqrt{s}=13$ TeV, 
  $\mathcal{E}(E_\text{obs})\sim \mathcal{O}(100)$ GeV.  
  Even if $b$ is as small as 1 GeV, the dependence is $1\%$.  
  When $k'$ is applied to leptons which have 
  their energy much higher than $\mathcal{E}(E_\text{obs})$, the bias on the corrected energy is approaching 
  $\mathcal{O}(1\%)$. When $k'$ is applied to leptons which have energy lower than $\mathcal{E}(E_\text{obs})$, 
  the bias is increasing fast and has no upper limit. This energy-dependent bias is caused by imperfect parameterization in Eq.~\ref{eq:wrongErelation}, and 
  will remain no matter how large are the data samples used in the calibration. 

   It is natural to try to introduce the offset term in the calibration as:
   
\begin{eqnarray}
  E'_\text{corr} = b' + k'\cdot E_\text{obs}.
\end{eqnarray}
  and we have 
\begin{eqnarray}\label{eq:multifactor_avg}
   b' + k' \cdot \mathcal{E}(E_\text{obs}) &=& b + k \cdot \mathcal{E}(E_\text{obs}) \nonumber \\
  k' - k &=& \frac{b - b'}{\mathcal{E}(E_\text{obs})}.
\end{eqnarray}
  Therefore the energy-dependent bias is:
  
\begin{eqnarray}\label{eq:uncontrol}
  \frac{E_\text{corr} - E_\text{truth}}{E_\text{obs}} &=& (k' - k) \cdot \left[ 1 - \frac{\mathcal{E}(E_\text{obs})}{E_\text{obs}} \right] \nonumber \\
    &=& \frac{b - b'}{\mathcal{E}(E_\text{obs})}\left[ 1- \frac{\mathcal{E}(E_\text{obs})}{E_\text{obs}} \right].
\end{eqnarray}

  $b'$ and $k'$ can be arbitrary because there is only one constraint from dilepton mass, 
  thus the dependence has no upper limit which seems even worse than the single scaling factor calibration. 
  To avoid such problem, $b$ factor 
  is always ignored in the calibration. 
  The dependence can be reduced if the 
  $b$ term itself is reduced with careful study from detector simulation. 
  However, 
  it is difficult and time-consuming. 

\subsection{I-C. New method}

  We present a new calibration method which allows to have both $b$ and $k$ parameters in the calibration function. 
  The new method first separates the dilepton events into subsamples with different 
  kinematic features, to introduce multiple mass constraints. Then, a technique to reduce the correlation between $b$ and $k$ 
  parameters is developed. As a result, the method can precisely determine values of $k$ and $b$ using 
  $Z\gamma^*\rightarrow \ell^+\ell^-$ events.
    
  To better explain the method and provide supporting tests, 
  $pp\rightarrow Z/\gamma^* \rightarrow \ell^+\ell^-$ events at $\sqrt{s}=13$ TeV are generated using the 
  {\sc pythia} generator~\cite{pythia}. 
  The total number of events in the sample is 72 M in full phase space, corresponding to a data sample of one year run of LHC. 
  To model the detector acceptance and the online threshold of lepton triggers at most 
  experiments, leptons are required to have their transverse momentum $p_T>25$ GeV.  A mass window cut of 
  $80 < M < 100$ GeV is also applied, as what usually has been done in real $Z/\gamma^*\rightarrow \ell^+\ell^-$ event selection. 

  The theory of this new method is described in section II. 
  In section III, the calibration is applied to the generator level sample, in which the 
  lepton energy is shifted by random energy scales, to estimate the uncertainty.
  In section IV, we discuss alternative 
  calibration methods for forward leptons and muons with charge-dependence. 
  Section V gives further discussion on mass mean obaservation, 
  and discussion on the correlation between energy scale and energy resolution. Section VI is a summary. 
  
\section{II. Theory of the New Method}\label{sec:theory}
\subsection{II-A. Multiple mass constraints}\label{sec:multimass}

  The new method is to apply $\eta$-dependent $k$ and $b$ parameterss to the observed lepton energy as correction:
  
\begin{eqnarray}
  E_\text{corr}(\eta) = b(\eta) + k(\eta) \cdot E_\text{obs}(\eta)
\end{eqnarray}

\begin{figure}[!hbt]
\begin{center}
 \epsfig{scale=0.4, file=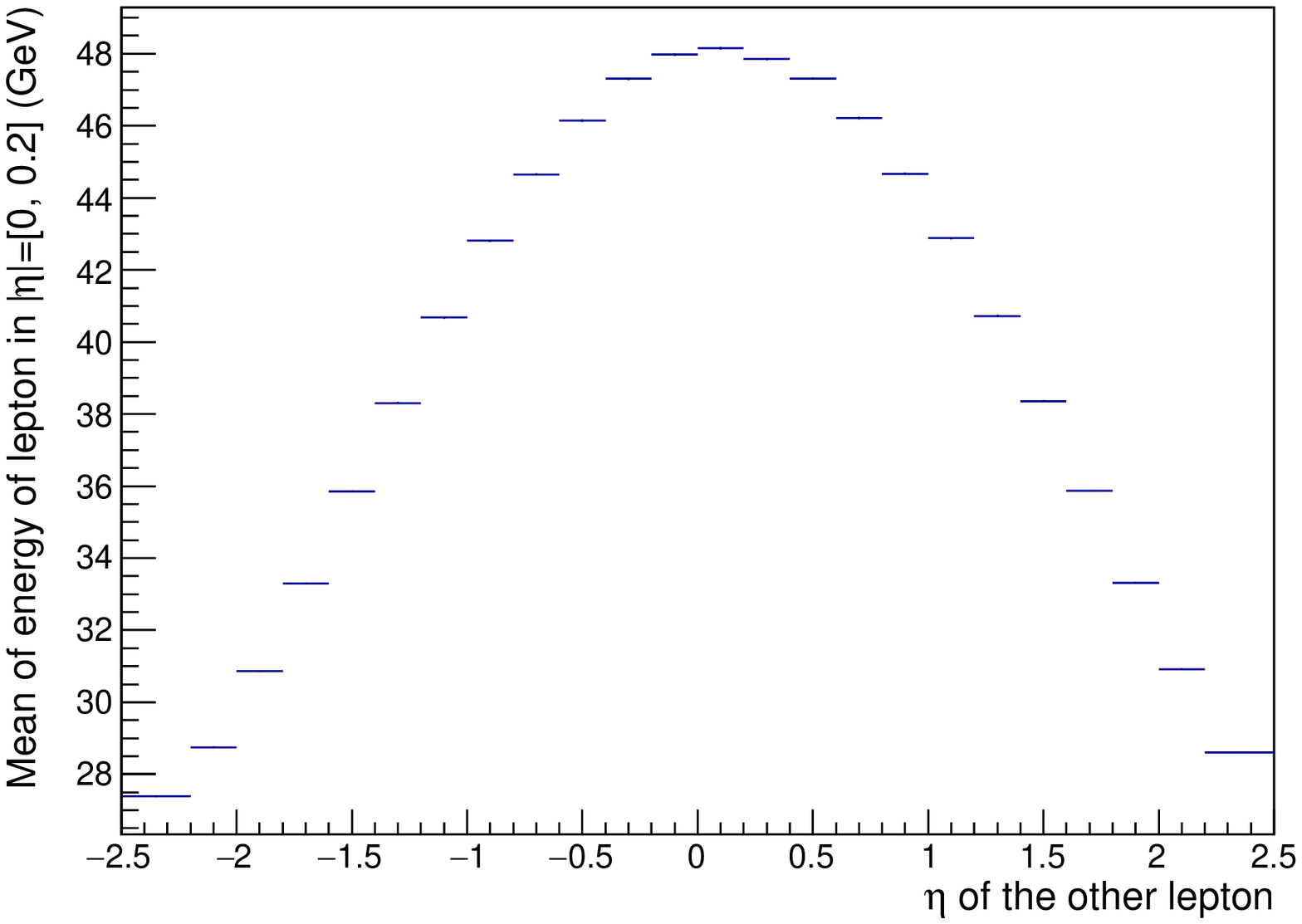}
 \epsfig{scale=0.4, file=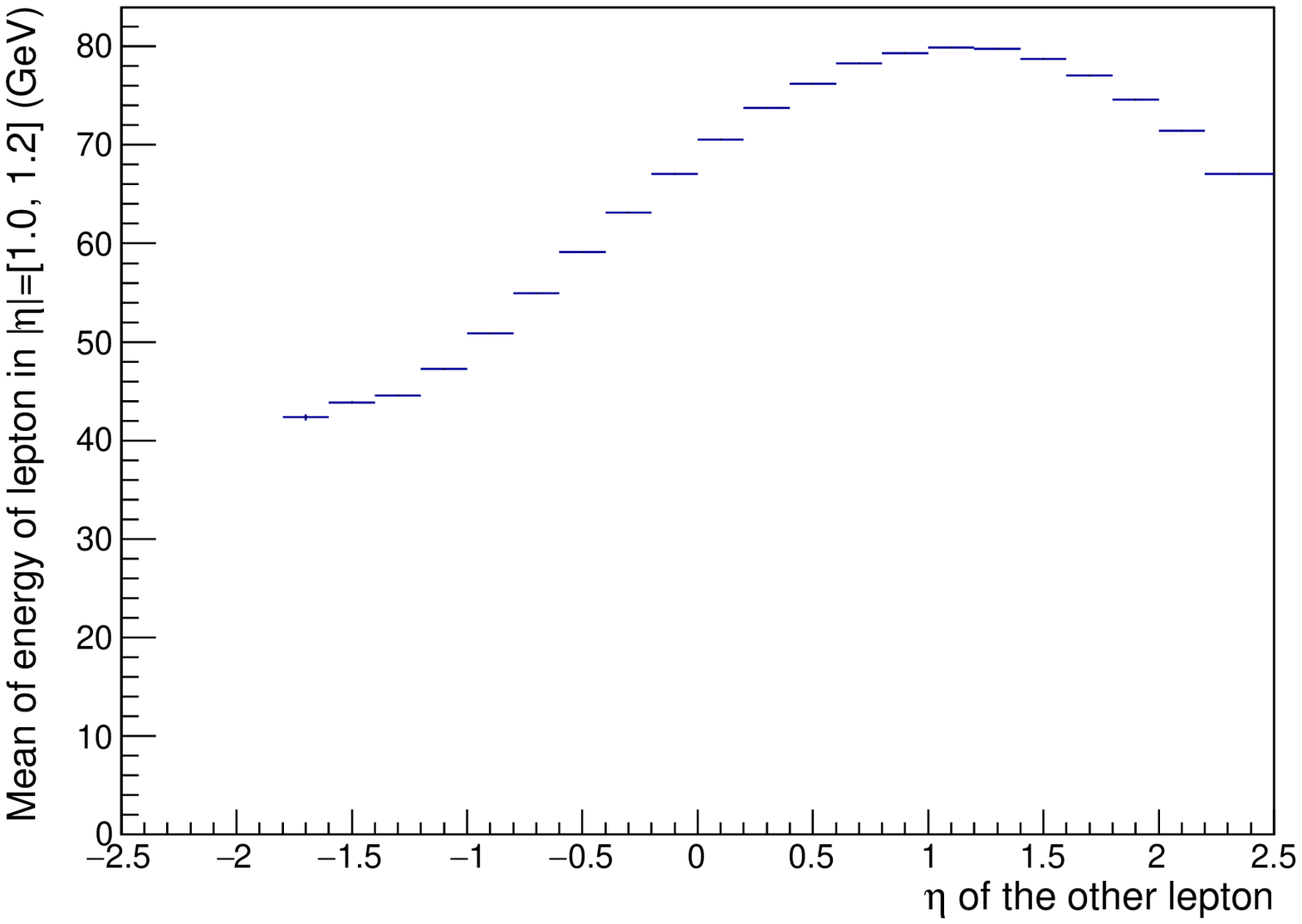}
 \epsfig{scale=0.4, file=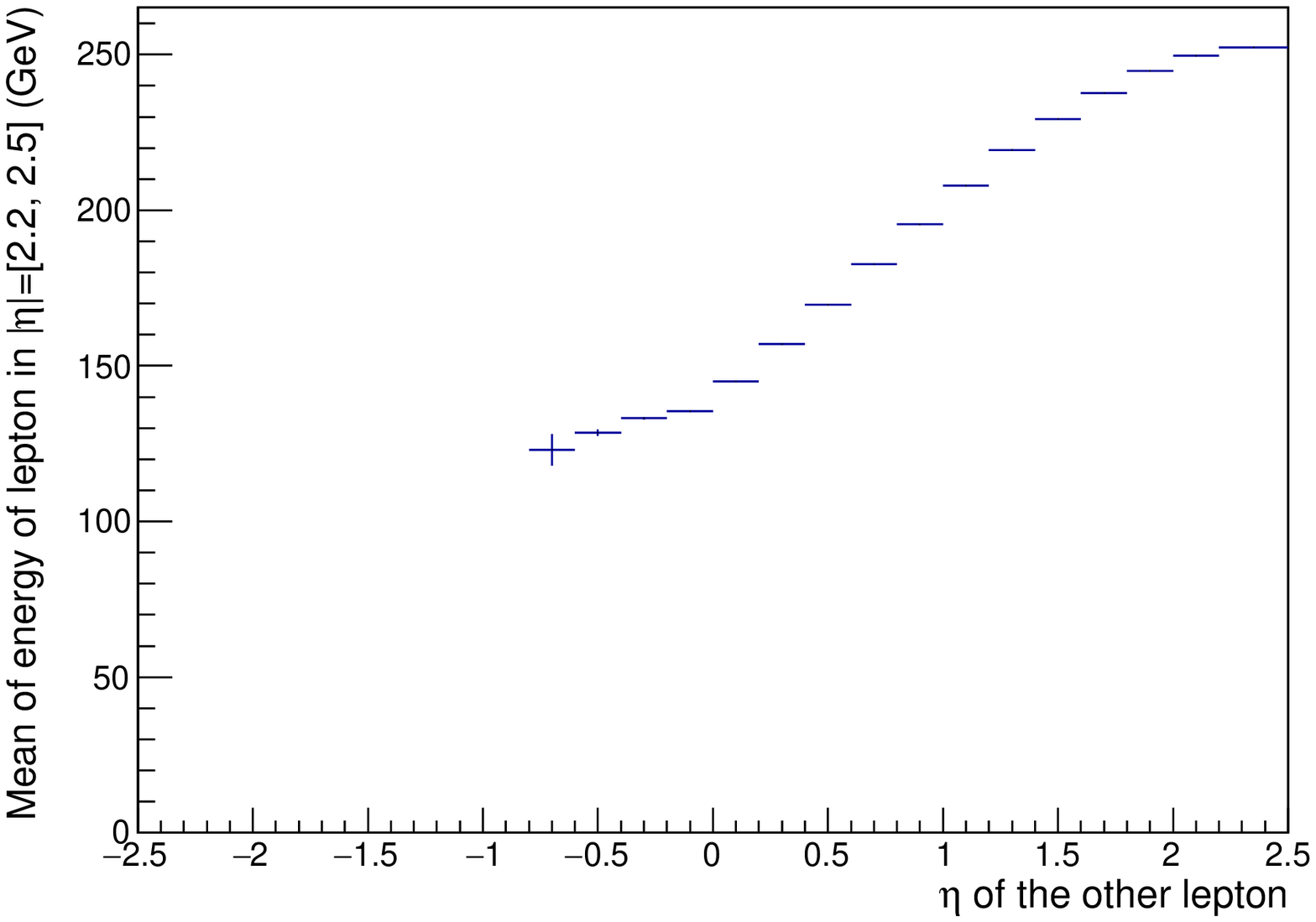}
\caption{\small $\mathcal{E}(E)$ in separated subsamples. The Y-axis is the energy mean of the 
lepton in a given $\eta$ region of three examples for [0, 0.2] (top), [1.0, 1.2] (middle) and [2.2, 2.5] (bottome). 
The X-axis is the $\eta$ 
of the other lepton in the dilepton events. Each plot shows the different energy of the leptons in the given $\eta$ region. 
Some bins are empty because there is no event if $\Delta \eta$ between 
two leptons is too large.}
\label{fig:kinematic}
\end{center}
\end{figure}

  To determine parameters of $k$ and $b$, the first step of the method has to introduce enough 
  physics constraints.
  It can be done by separating the $Z/\gamma^* \rightarrow \ell^+\ell^-$ events into subsamples 
  based on the opening angle between leptons. 
  At hadron collider, leptons from heavy boosted $Z$ boson decay are close 
  to each other, and have higher energy and smaller opening angle, while leptons from almost stationary $Z$ boson decay tend to have 
  lower energy and larger opening angle. 
  Thus, the dilepton events can be separated into subsamples with different lepton energies by cutting on the opening 
  angle. 
  Since cutting on the opening angle is separating more on the $Z$ boson boost, 
  the line shapes of dilepton mass spectrum are still good.
  As a result, in each 
  separated subsample, the constraint from mean value of the dilepton mass, $\mathcal{E}(M)$, corresponds to a unique value of 
  the mean value of lepton energy $\mathcal{E}(E_\text{obs})$, providing several points when 
  determining the line in Eq.~\ref{eq:linearErelation}.  
  Fig.~\ref{fig:kinematic} shows an example for the above description. 
  Subsamples are made to contain at least one lepton in a given $\eta$ region, but have the other lepton in different $\eta$ 
  regions, so that the leptons in the given $\eta$ region have different energies.

  For each $\eta$ region, there are two parameters, $k(\eta)$ and $b(\eta)$, to be determined. Thus, with a total number 
  $N$ of $\eta$ regions involved in a dilepton sample, there are $2N$ factors in total, while the number of 
  subsamples is $N(N - 1)$. 
  To have 
  enough mass constraints, we must have $N(N -1) \ge 2N$. Therefore, the minimal 
  value of $N$ is three, indicating that 6 $k(\eta)$ and $b(\eta)$ factors have to be 
  determined together as a group with six subsamples.

\begin{figure}[!hbt]
\begin{center}
\includegraphics[width=0.4\textwidth]{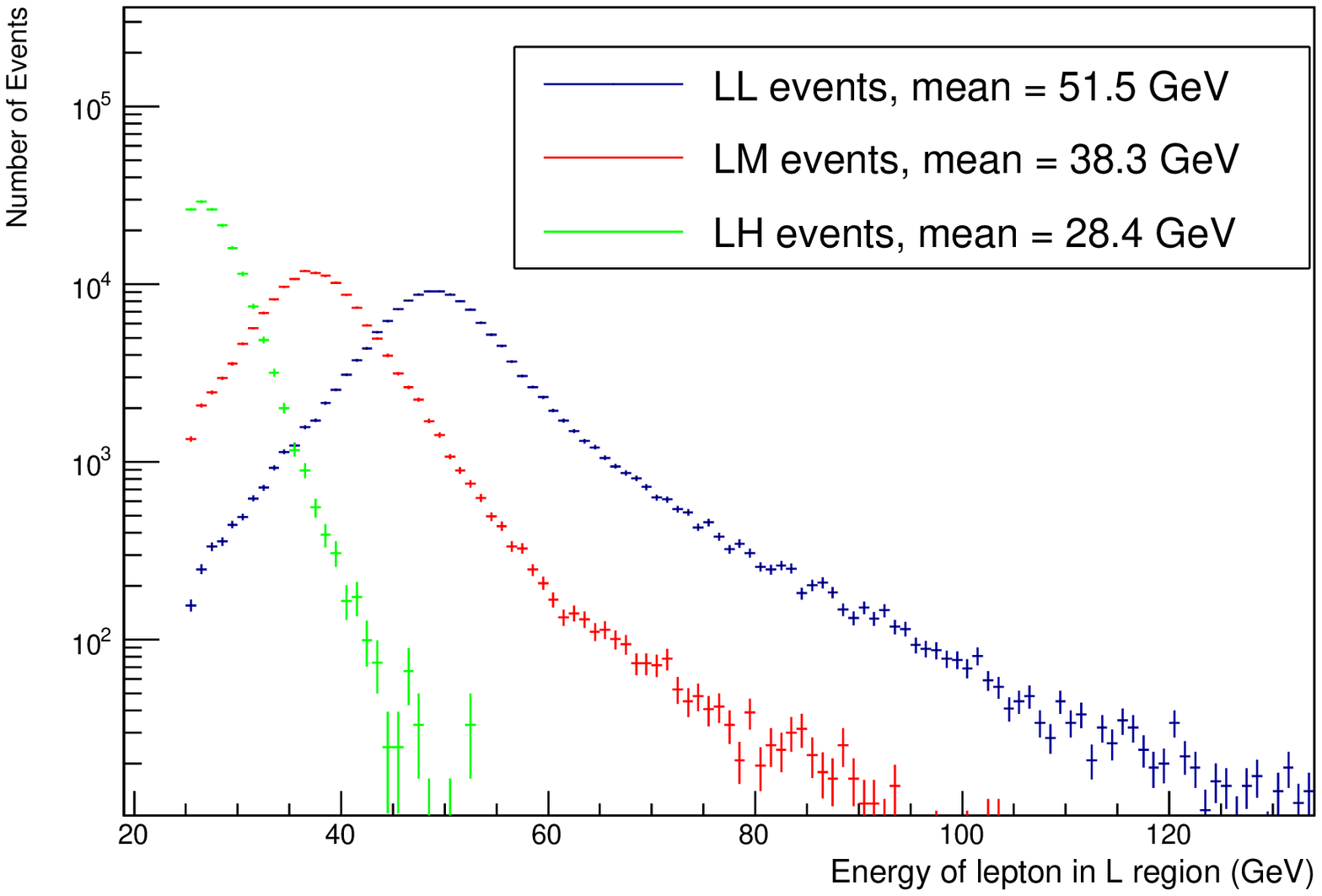}
\includegraphics[width=0.4\textwidth]{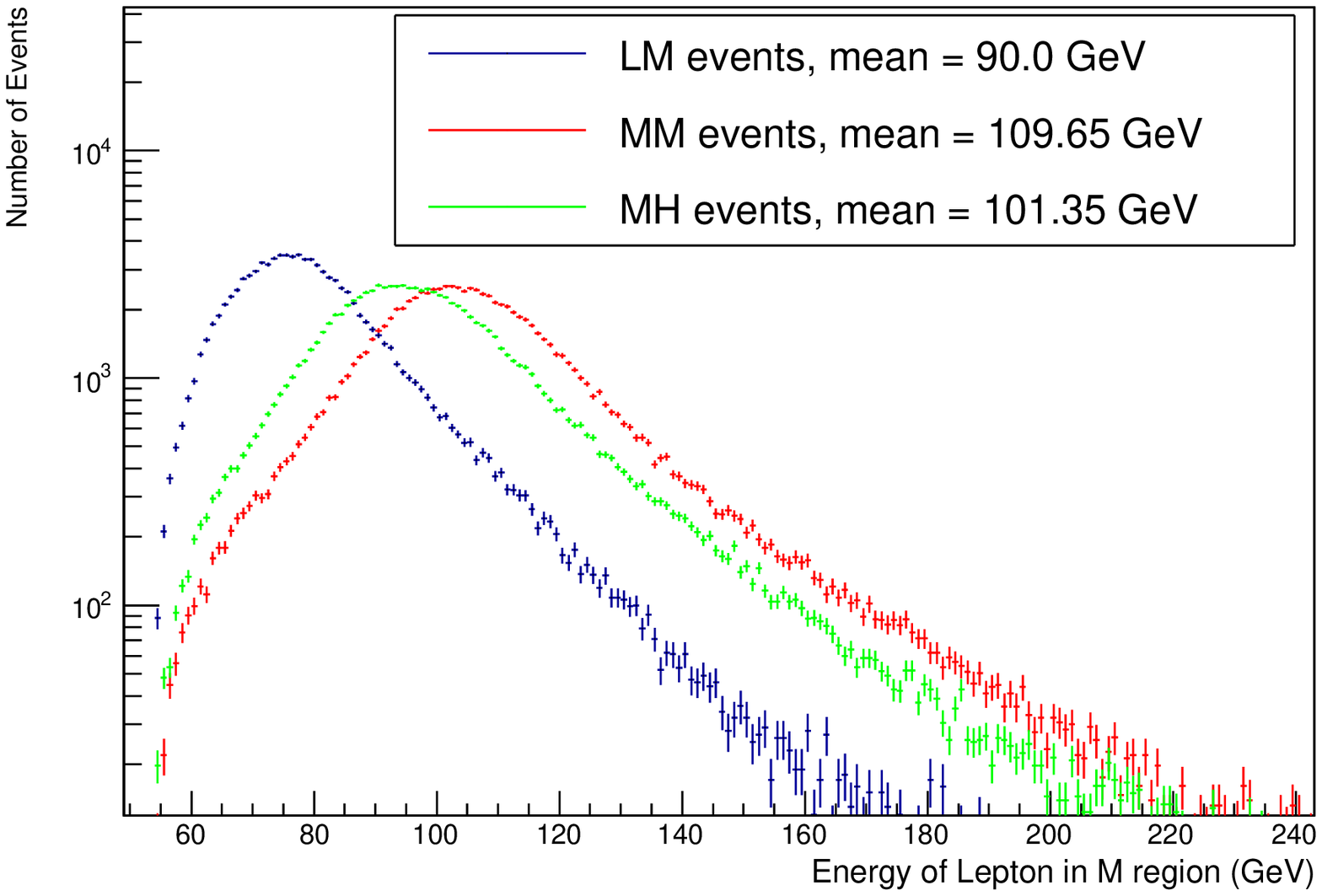}
\includegraphics[width=0.4\textwidth]{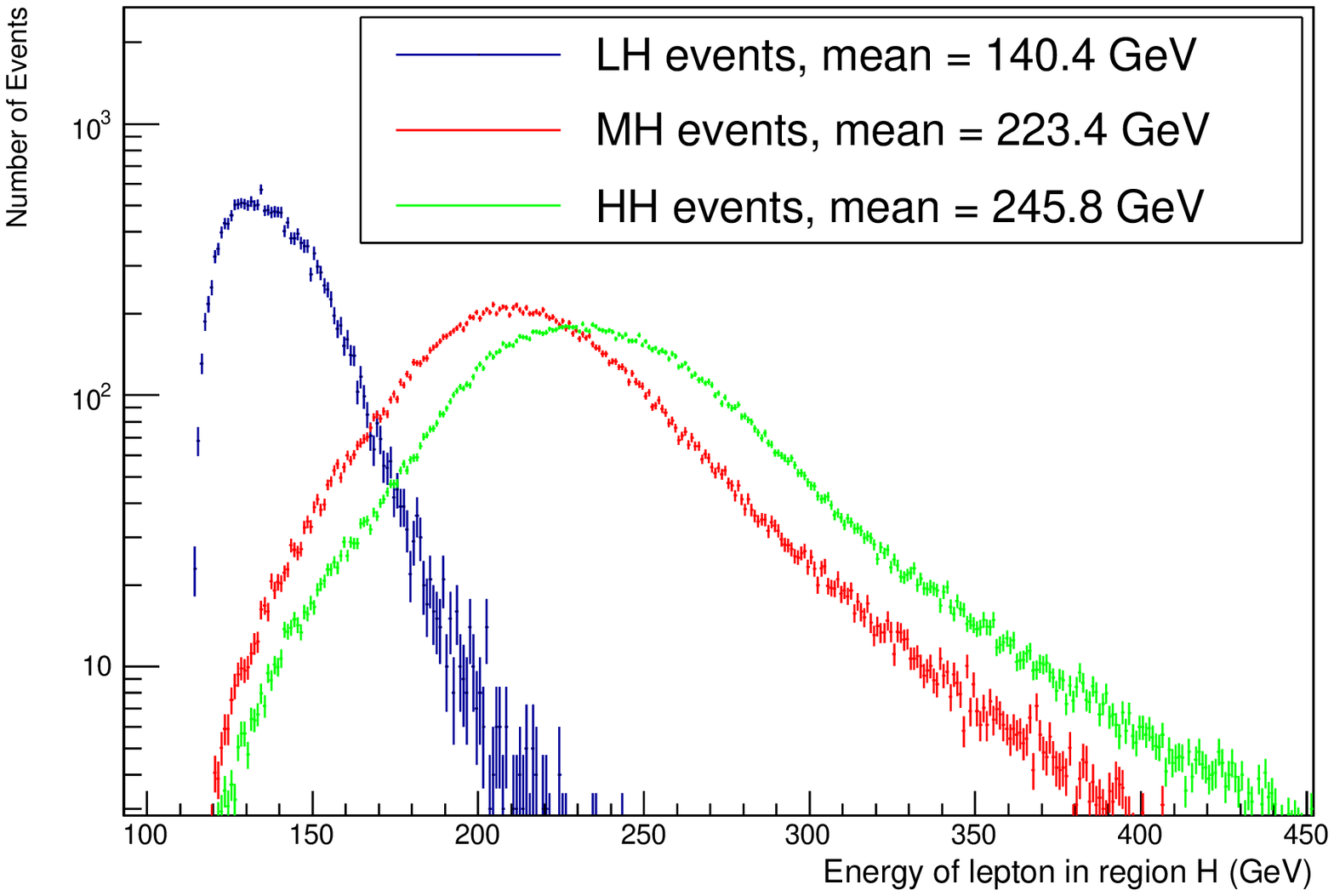}
\caption{\small Energy spectrum of the lepton in $Z\rightarrow \ell^+\ell^-$ events. 
The X-axis is the energy of the lepton in $L=[0, 0.2]$ (top), $M=[-1.6, -1.4]$ (middle) and $H=[-2.5,-2.2]$ (bottom) 
region. The three colors corresponds to the $\eta$ region of the other lepton. Spectrum in one plot are scaled to the 
same normalization.}
\label{fig:Epoints}
\end{center}
\end{figure}

  We introduce a group of three $\eta$ regions denoted as $L$, $M$ and $H$. The six subsamples, according to the combination 
  of lepton $\eta$ is $LL$, $LM$, $LH$, $MM$, $MH$ and $HH$ (e.g. $LL$ means both lepton in $L$ region, $LH$ 
  means one lepton in $L$ and the other in $H$ region). For parameters related to the $L$ region, $k_L$ and $b_L$, 
  $LL$, $LM$ and $LH$ events provide three constraints. For parameters related to the $M$ region, $k_M$ and $b_M$, 
  $LM$, $MM$ and $MH$ events provide three constraints. For parameters related to the $H$ region, $k_H$ and $b_H$, 
  $LH$, $MH$ and $HH$ events provide three constraints.
  The constraints can be expressed as six equations:
  
\begin{eqnarray}\label{eq:massnarray}
  M^2_\text{true}[\alpha \beta] &=& M^2_\text{corr}[\alpha \beta] \nonumber \\
   &=& 2(b_\alpha + k_\alpha \cdot E_\text{obs}[\alpha])\times \nonumber \\
      & & (b_\beta + k_\beta \cdot E_\text{obs}[\beta ])(1-\cos\theta)
\end{eqnarray}
  where $\alpha, \beta = L, M, H$ ($LM$ and $ML$ are same, $LH$ and $HL$ are same, and $MH$ and $HM$ are same). 
  After calculating the mean, 
  the above constraints are equivalent to the following nine equations:

\begin{eqnarray}\label{eq:fulleqnarray}
  b_\alpha + k_\alpha \cdot \mathcal{E}(E^\alpha_{\alpha\beta}[\text{obs}]) &=& \mathcal{E}(E^\alpha_{\alpha \beta}[\text{true}])
\end{eqnarray}
  where $E^\alpha_{\alpha \beta}[\text{corr}]$ and $E^\alpha_{\alpha \beta}[\text{true}]$ ($\alpha, \beta = L, M, H$) 
  are the energy after correction and its corresponding true value of the leptons appears in region 
  $\alpha$ while the other lepton appears in region $\beta$. 
  Since we only have six mass constraints, the nine equations in Eq.~\ref{eq:fulleqnarray} are 
  not independent of each other.  
  Fig.~\ref{fig:Epoints} shows an example of designed $\eta$ regions, $L = [0, 0.2]$, $M=[-1.6, -1.4]$ and $H=[-2.5, -2.2]$, 
  and the different lepton energies in subsamples.

\begin{figure}[!hbt]
\begin{center}
\includegraphics[width=0.45\textwidth]{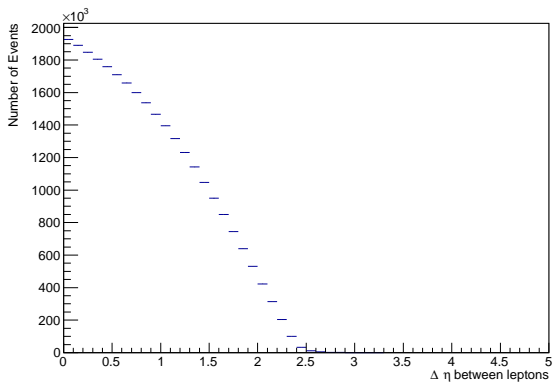}
\caption{\small Distribution of $| \Delta \eta |$
between leptons in $Z/\gamma^* \rightarrow \ell^+\ell^-$ events.}
\label{fig:deltaeta}
\end{center}
\end{figure}

  The more significant $\mathcal{E}(E^\alpha_{\alpha \beta})$ ($\alpha, \beta = L, M, H$) differ 
  from each other, the higher sensitivity we have in the determination of $k$ and $b$ parameters. 
  To make $\mathcal{E}(E^\alpha_{\alpha \beta})$ significantly differ, $L$, $M$ and $H$ regions 
  should have large $\Delta \eta$ between them. But it reduces the statistics in subsamples, as shown 
  in Fig.~\ref{fig:deltaeta}. At LHC, difference between $\mathcal{E}(E^\alpha_{\alpha \beta})$ 
  is more important than statistics, because the data sample of dilepton events is large enough.

\subsection{II-B. Correlation}\label{sec:correlation}
  
  Eq.~\ref{eq:massnarray} in principle provides enough constraints. A naive try is 
  to determine $k$ and $b$ values by minimizing  
  a $\chi^2$ defined as:
  
\begin{eqnarray}\label{eq:definechi2}
  \chi^2 = \sum_{\alpha\beta} \frac{\left[ \mathcal{E}(M_\text{corr}[\alpha\beta]) - \mathcal{E}(M_\text{true}[\alpha\beta]) \right]^2}{\sigma^2_M[\alpha\beta]}
\end{eqnarray}
  where $\alpha, \beta = L, M, H$ refer to the six event categories. $\mathcal{E}(M_\text{corr}[\alpha\beta])$ is the mean 
  of the mass spectrum after calibration in each event category. $\mathcal{E}(M_\text{true}[\alpha\beta])$ is 
  the reference mean of the mass spectrum. $\sigma_M[\alpha\beta]$ is the uncertainty of the 
  observed mass mean, which is statistically dominated by the data subsamples. 
  
  However a fit always stops at a very large $\chi^2$ value and gives unreasonable values for 
  $k$ and $b$. It is simply caused by correlation between parameters, especially the correlation between $k$ and 
  $b$ in the same $\eta$ region. The correlation gives many extreme points (but not minimal points) of $\chi^2$ value 
  in the $k$-$b$ six dimension space, while only one of them is the physics solution of 
  $k$ and $b$ values. In another word, it is difficult to provide a reasonable range for the unknown factors in the fit. 
  This difficulty could be solved  
  by using some advanced fit method, 
  but is time-consuming. As data accumulates very 
  fast at hadron colliders, we need to have a procedure that can reduce the correlation easily and fast. 
  
\subsection{II-C. Reduce the correlation}\label{sec:reduceCorr}

  The idea to reduce the correlation is to construct a relationship between $k$ and $b$ parameters 
  in each $\eta$ region. In this relationship, quantities except $k$ and $b$ should be 
  known, so that $b$ can be expressed using 
  $k$. Eq.~\ref{eq:singlefactor_avg} is a good choice. But as discussed, it holds as an 
  approximation. Here we do a better derivation.

  For events with both leptons in the same $\eta$ region, the mass can be re-written as:

\begin{footnotesize}  
\begin{eqnarray}
 M^2_\text{true} &=& 2 (b + kE^\text{obs}_1)(b+k E^\text{obs}_2) (1-\cos\theta_{12}) \nonumber \\
   &=& \left( k + \frac{b}{E^\text{obs}_1}\right) \cdot \left( k + \frac{b}{E^\text{obs}_2}\right)\cdot M^2_\text{obs}
\end{eqnarray}
\end{footnotesize}

  Because $\mathcal{E}(X\cdot Y) = \mathcal{E}(X)\cdot \mathcal{E}(Y) + cov(X, Y)$ and 
  $\mathcal{E}(X^2) = \mathcal{E}(X)^2 + \mathcal{D}(X)$ (where $cov$ means co-variance 
  and $\mathcal{D}$ means variance),
  we have 
   
\begin{footnotesize}  
\begin{eqnarray} \label{eq:fullexpension}
 \mathcal{E}(M^2_\text{true}) &=& \mathcal{E}(M_\text{true})^2 + \mathcal{D}(M_\text{true}), \nonumber \\
 \mathcal{E}(M^2_\text{true}) &=& \mathcal{E}\left[ \left( k + \frac{b}{E^\text{obs}_1} \right) \cdot \left( k + \frac{b}{E^\text{obs}_2} \right) \cdot M^2_\text{obs}\right] \nonumber \\
   &=& \mathcal{E}\left[ \left( k+\frac{b}{E^\text{obs}_1}\right) \left( k+\frac{b}{E^\text{obs}_2}\right) \right] \cdot  \mathcal{E}(M_\text{obs})^2 \nonumber \\
   & & + \mathcal{E}\left[ \left( k+\frac{b}{E^\text{obs}_1}\right) \left( k+\frac{b}{E^\text{obs}_2}\right) \right] \cdot  \mathcal{D}(M_\text{obs}) \nonumber \\
   & & + cov\left[ \left( k + \frac{b}{E^\text{obs}_1} \right) \left( k + \frac{b}{E^\text{obs}_2} \right), M^2_\text{obs}\right].
\end{eqnarray}
\end{footnotesize}
  When $b=0$, Eq.~\ref{eq:fullexpension} leads to a conclusion that ratio of mass is consistent with ratio of 
  energy: 

$$
 \frac{\mathcal{E}(E_\text{true})}{\mathcal{E}(E_\text{obs})} = \frac{\mathcal{E}(M_\text{true})}{\mathcal{E}(M_\text{obs})}.
$$
  When $b \ne 0$, a correction $\epsilon$ has to be added due to the co-variance between $M$, $E_1$ and $E_2$:
  
\begin{eqnarray}\label{eq:bkrelation}
  \frac{\mathcal{E}(E_\text{true})}{\mathcal{E}(E_\text{obs})} &=& \frac{\mathcal{E}(M_\text{true})}{\mathcal{E}(M_\text{obs})} + \epsilon \nonumber \\
  \mathcal{E}(E_\text{true}) &=& b + k\cdot \mathcal{E}(E_\text{obs}) \nonumber \\
    &=& \mathcal{E}(E_\text{obs})\cdot  \left[ \frac{\mathcal{E}(M_\text{true})}{\mathcal{E}(M_\text{obs})} + \epsilon \right].
\end{eqnarray}

\begin{figure}[!hbt]
\begin{center}
\includegraphics[width=0.4\textwidth]{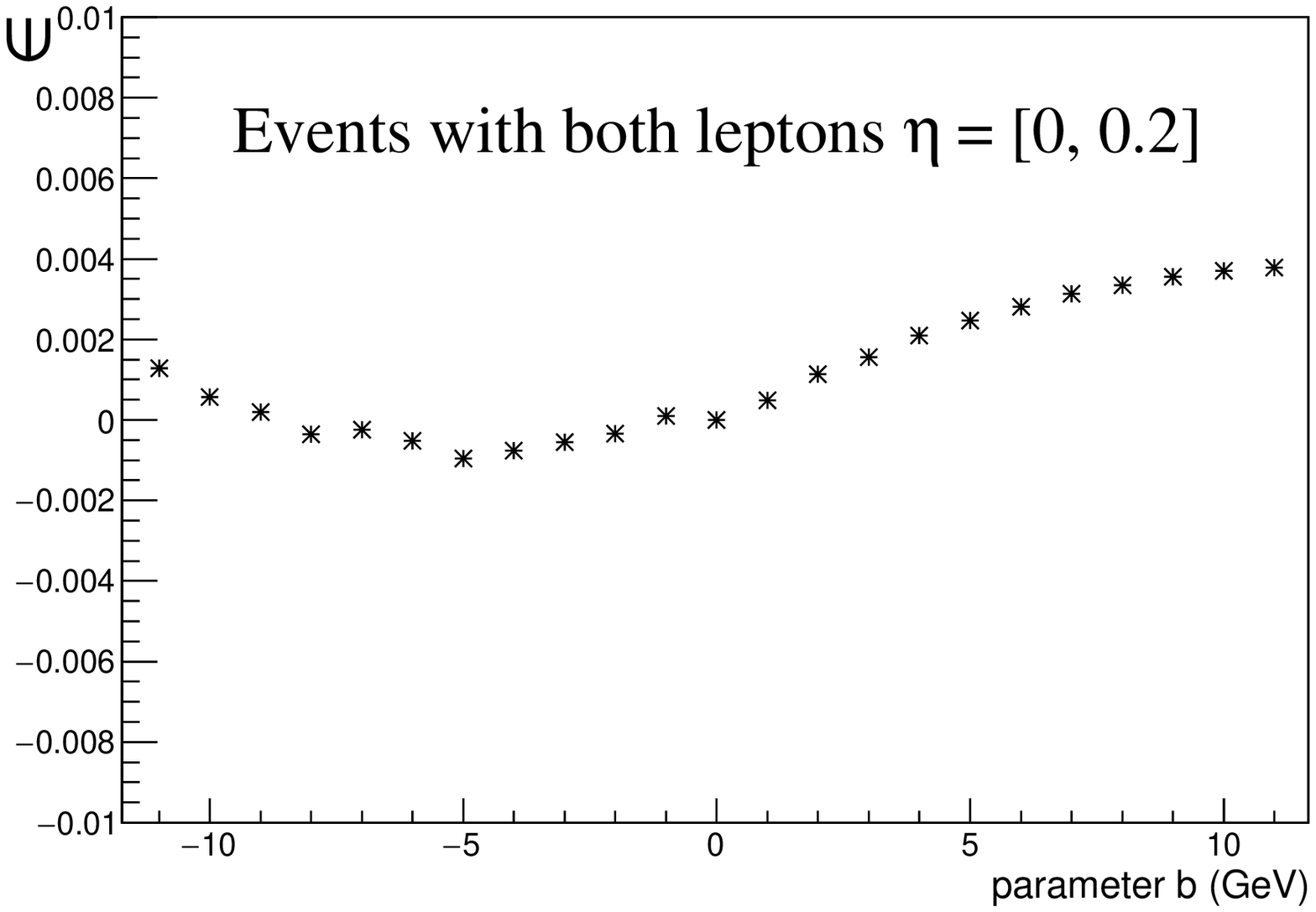}
\includegraphics[width=0.4\textwidth]{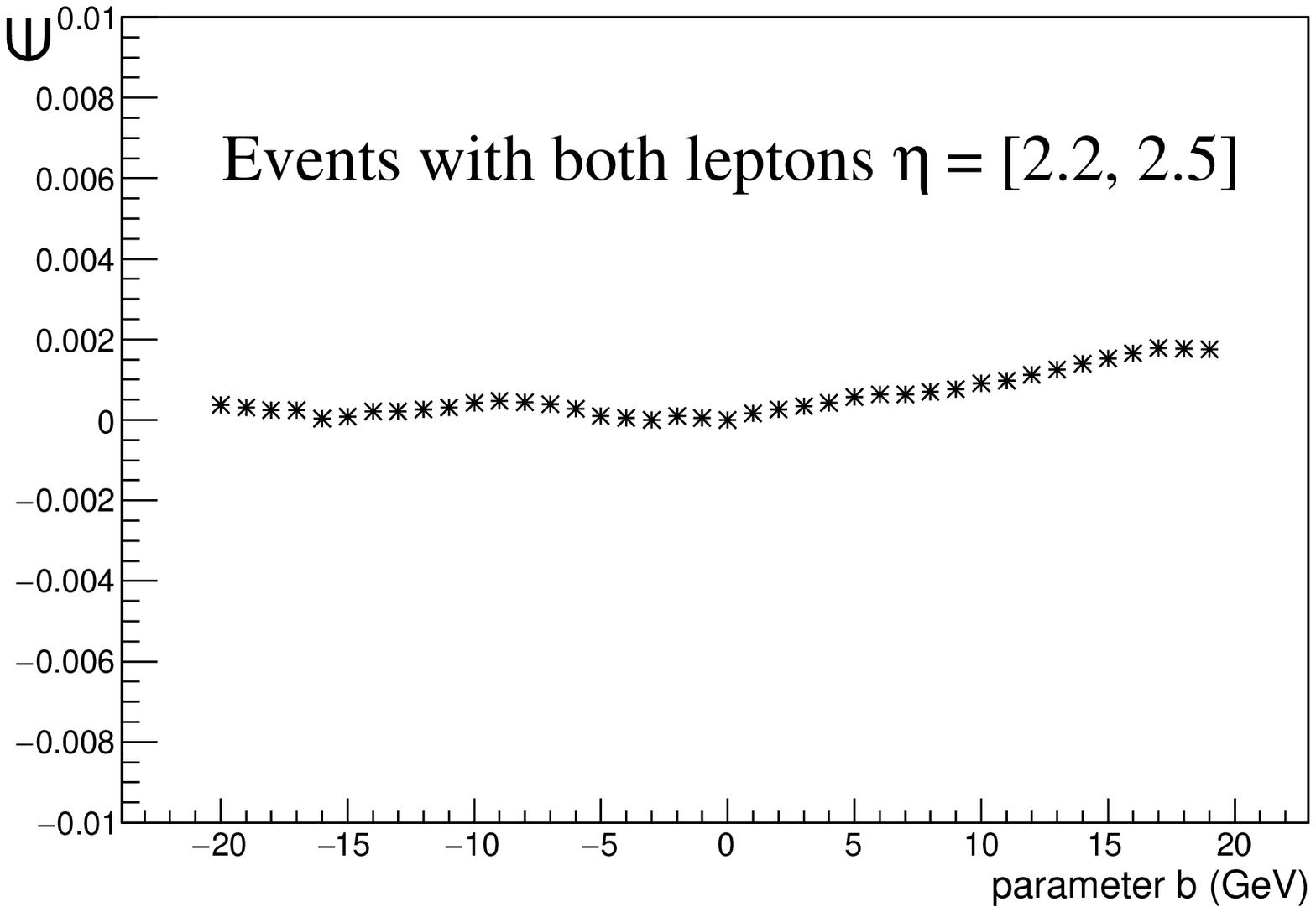}
\caption{\small Estimated value of $\epsilon$ as a function of $b$ in events with both leptons 
$\eta = [0, 0.2]$ (top) and $\eta = [2.2, 2.5]$ (bottom). In the top plot, lepton energy is around 50 GeV. 
When $b$ is changing in a range of $\pm 10$ GeV, the fitting range for $b/E \sim \pm 0.2$ while 
$|\epsilon |< \pm 0.002$. In the bottom plot, lepton energy is around 200 GeV. 
When $b$ is changing in a range of $\pm 20$ GeV, the fitting range for $b/E \sim \pm 0.1$ while $|\epsilon | <0.001$.}
\label{fig:epsilon}
\end{center}
\end{figure}

  Note that all calculations of $\mathcal{E}$ are based on an ensemble of events where 
  both leptons are in the same $\eta$ region. 
  $b$ can be expressed using Eq.~\ref{eq:bkrelation}. $\mathcal{E}(E_\text{obs})$, 
  $\mathcal{E}(M_\text{obs})$ and $\mathcal{E}(M_\text{true})$ can be easily observed or known. $\epsilon$ is 
  in principle also known. However calculating co-variance between energy and mass spectrum is 
  too troublesome. Instead, we can leave $\epsilon$ as a free parameter in the final fit. Fitting for $k$ and $\epsilon$ 
  is much easier than fitting for $k$ and $b$, because with mass constraints used in Eq.~\ref{eq:bkrelation}, 
  $\epsilon$ is reduced to a very small value.  
  The estimated values of $\epsilon$ as a function of $b$ are shown in Fig.~\ref{fig:epsilon}.
  $\epsilon$ is 10 times smaller than $b/E$. 
  So the correlation between $k$ and $\epsilon$ causes no trouble in the fit because we can 
  give $\epsilon$ a fit range much smaller than the range for $b$.

  One may say, why not observe $\mathcal{E}(E_\text{true})$ and $\mathcal{E}(E_\text{obs})$, so that 
  the relationship between $b$ and $k$ is directly provided as the first row in Eq.~\ref{eq:bkrelation}? 
  As discussed in section I-A, difference between $\mathcal{E}(E_\text{true})$ and $\mathcal{E}(E_\text{obs})$ 
  is determined not only by energy scale, but also PDFs and QCD calculation. 
  An example is shown in Tab.~\ref{tab:pdftest}. Two samples of $pp\rightarrow Z/\gamma^* \rightarrow \ell^+\ell^-$ 
  events are generated with different PDF sets randomly chosen from the NNPDF~\cite{NNPDF}. 
  The mean of lepton energy differs by $3\%$, while the mean of dilepton 
  mass only differs by $0.006\%$. 
  Therefore, we do not 
  observe $\mathcal{E}(E_\text{true})$ and $\mathcal{E}(E_\text{obs})$ simultaneously. Instead, any comparison 
  between observation and its corresponding truth value should be done with mass. 
 
\begin{table}[h]
\begin{center}
\begin{tabular}{l|c|c}
\hline \hline
      PDF &   Mean of & Mean of \\
       & dilepton mass & lepton energy \\
\hline
 NNPDF3.1 & 90.693 GeV & 246.199 GeV \\
 NLO No. 397  & & \\
\hline
 NNPDF3.1 & 90.688 GeV & 254.442 GeV \\
 LO No. 36 & & \\
\hline \hline
\end{tabular}
\caption{\small Mean of dilepton mass and lepton energy in $p\bar{p}\rightarrow Z/\gamma^* \rightarrow \ell^+\ell^-$ 
  events generated using two different PDF sets. Events are generated in mass range from 60 to 130 GeV.}
\label{tab:pdftest}
\end{center}
\end{table}

\subsection{II-D. Fit procedure}

  The fit procedure is described in this section. 
  From the relationship between $k$ and $b$ parameters observed using $LL$, $MM$ and 
  $HH$ events, we have

\begin{footnotesize}  
\begin{eqnarray}
  b_L &=& -\mathcal{E}(E^L_{LL}[\text{obs}]) \cdot k_L  \nonumber \\
     & & + \mathcal{E}(E^L_{LL}[\text{obs}])\cdot \left[ \frac{\mathcal{E}(M_{LL}[\text{true}])}{\mathcal{E}(M_{LL}[\text{obs}])} +\epsilon_L \right] \nonumber \\
  b_M &=& -\mathcal{E}(E^M_{MM}[\text{obs}]) \cdot k_M \nonumber \\
    & & + \mathcal{E}(E^M_{MM}[\text{obs}])\cdot \left[ \frac{\mathcal{E}(M_{MM}[\text{true}])}{\mathcal{E}(M_{MM}[\text{obs}])} +\epsilon_M \right] \nonumber \\
  b_H &=& -\mathcal{E}(E^H_{HH}[\text{obs}]) \cdot k_H \nonumber \\
    & & + \mathcal{E}(E^H_{HH}[\text{obs}])\cdot \left[ \frac{\mathcal{E}(M_{HH}[\text{true}])}{\mathcal{E}(M_{HH}[\text{obs}])} +\epsilon_H \right].
\end{eqnarray}
\end{footnotesize}

  Then, values of $k_L$, $k_M$, $k_H$, $\epsilon_L$, $\epsilon_M$ and $\epsilon_H$ can be determined 
  by a fit minimizing the $\chi^2$ defined as in Eq.~\ref{eq:definechi2}. As discussed in section II-C, the fit range for 
  $\epsilon$ is much narrower than that for $b$.

\section{III. Generator level test}\label{sec:tests}

  The method is tested at the generator level. A $pp\rightarrow Z/\gamma^* \rightarrow \ell^+\ell^-$ sample 
  is generated with statistics equivalent to the integrated luminosity of $\sim 35$ fb$^{-1}$ at 13 TeV 
  LHC. 
  As described in section II, the calibration is determining $k$ and $b$ in three $\eta$ regions as 
  a group. Table~\ref{tab:bins} is an example of bins for $\eta = [-2.5, 2.5]$. 

\begin{table}[h]
\begin{center}
\begin{tabular}{l|c|c|c|}
\hline \hline
      Group &   $H$ region & $M$ region & $L$ region \\
\hline
 1  & [-2.5, -2.2] & [-1.6, -1.4] & [0, 0.2] \\
\hline
 2  & [-2.2, -2.0] & [-1.4, -1.2]  & [0.2, 0.4] \\
\hline
 3  & [-2.0, -1.8] & [-1.2, -1.0] & [0.4, 0.6] \\
\hline
 4  & [-1.8, -1.6] & [-1.0, -0.8] & [0.6, 0.8] \\
\hline
 5  & [2.2, 2.5] & [1.4, 1.6] & [-0.2, 0] \\
\hline
 6  & [2.0, 2.2] & [1.2, 1.4] & [-0.4, -0.2] \\
\hline
 7  & [1.8, 2.0] & [1.0, 1.2] & [-0.6, -0.4] \\
\hline
 8  & [1.6, 1.8] & [0.8, 1.0] & [-0.8, -0.6] \\
\hline \hline
\end{tabular}
\caption{\small An example of binning in the calibration procedure. 24 regions in 8 groups fully cover the 
$\eta$ region from $-2.5$ to $+2.5$.}
\label{tab:bins}
\end{center}
\end{table}

  For each $\eta$ region, the energy of the lepton is shifted by $k$ and $b$ parameters:

\begin{eqnarray}
  E_\text{shift}(\eta) = k_\eta \cdot E(\eta) + b_\eta.
\end{eqnarray}
  Values of $k_\eta$ are given by a uniform distribution in a region of 0.97 to 1.03. Values of $b_\eta$ are  
  given by a uniform distribution in a region of $-3$ to $+3$ GeV.
  Then, a $p_T>25$ GeV cut 
  is applied to the lepton. 
  A nominal $pp\rightarrow Z/\gamma^* \rightarrow \ell^+\ell^-$ sample 
  is also generated without any energy shift.  The calibration method described in section II is applied 
  to the energy shifted sample to match with the nominal sample. 
  
\begin{figure}[h]
\begin{center}
\includegraphics[width=0.4\textwidth]{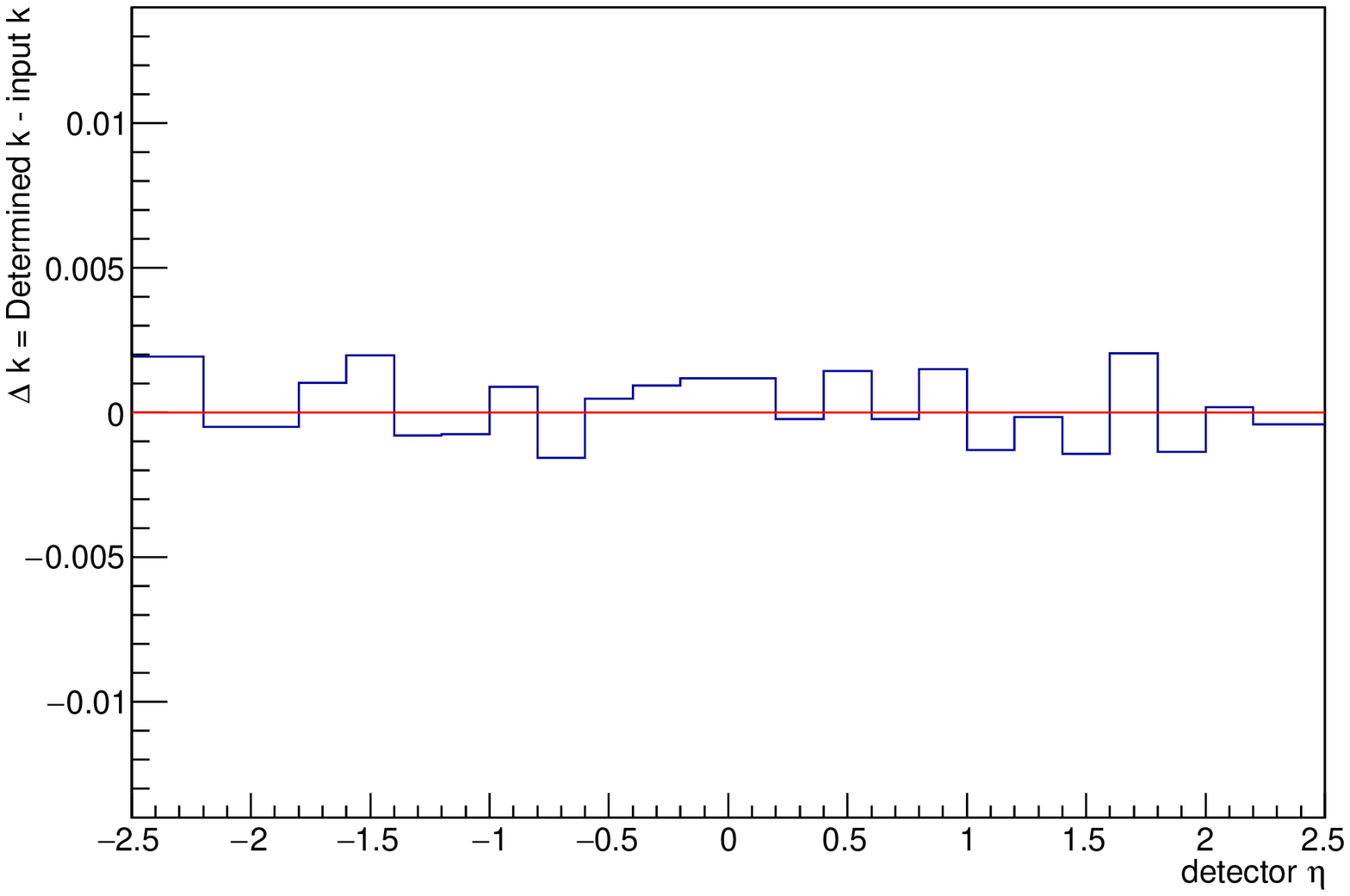}
\includegraphics[width=0.4\textwidth]{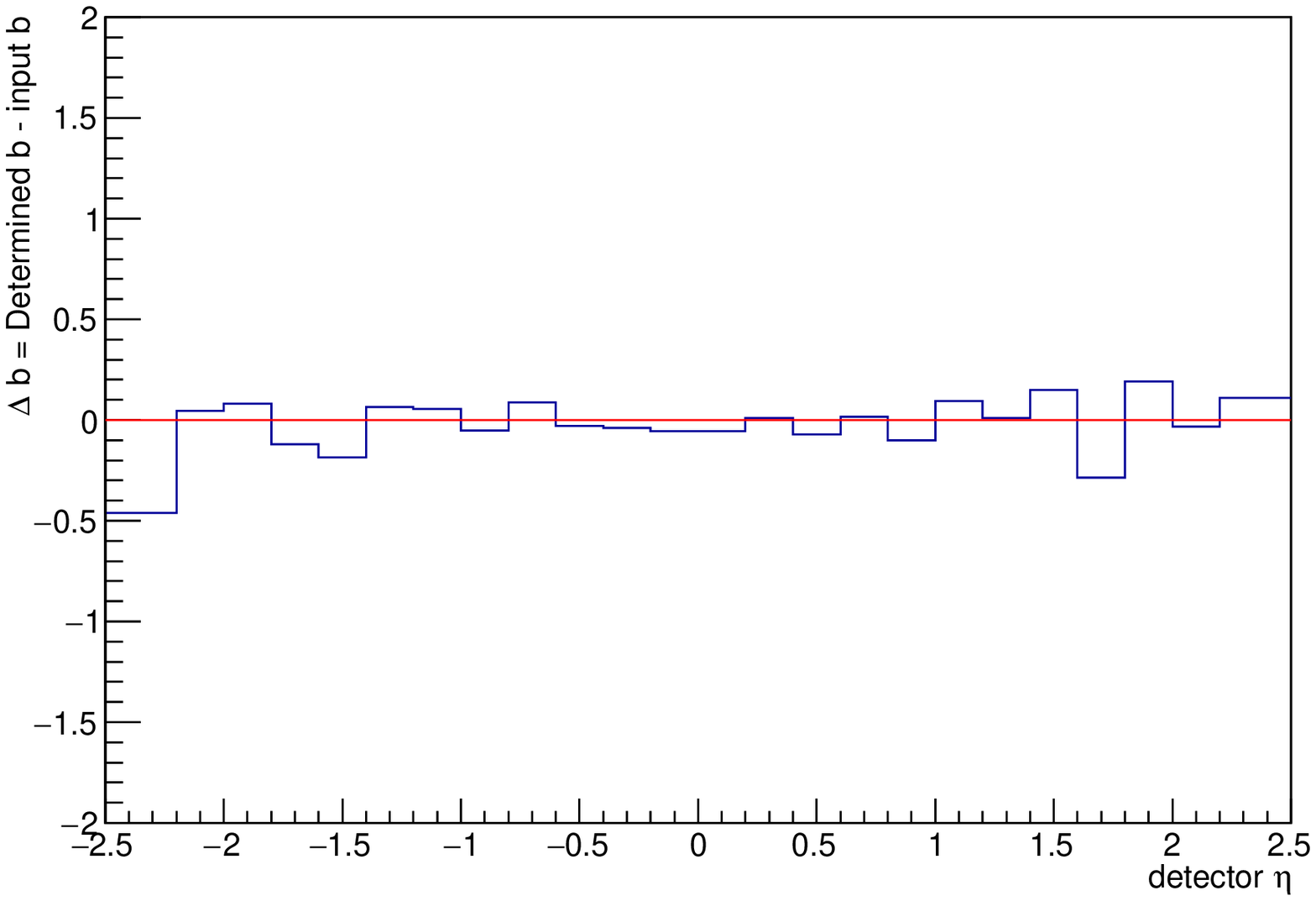}
\caption{\small $\delta k$ and $\delta b$ between the input values and the fitted values. relative uncertainty is 
smaller than 0.002. $\delta b$ is in measurement of GeV.}
\label{fig:deltaPara}
\end{center}
\end{figure}

  Fig.~\ref{fig:deltaPara} shows the difference between the input values of $k_\eta$ and $b_\eta$
  and their fitted values using the calibration procedure. As we can see, $\delta k$ and $\delta b$ 
  are very small, randomly located around 0. The relative uncertainties in $\delta k$ and $\delta b$ are 
  smaller than 0.002, which is dominated by the statistics of the two samples.

\begin{figure}[h]
\begin{center}

\includegraphics[width=0.45\textwidth]{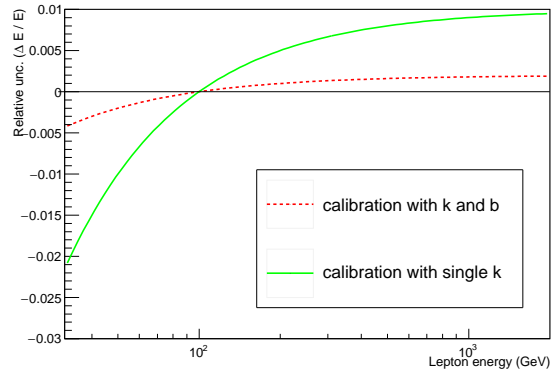}
\caption{\small Relative uncertainty $\Delta E / E$ as a function of $E$. For calibration with both $k$ and $b$ factors, 
running speed is set to $\delta k = 0.002$. For calibration with single scaling $k$ factor, the ignored factor is set 
to $b = 1$ GeV.}
\label{fig:unc}
\end{center}
\end{figure}

  The uncertainty of the calibration can be estimated using Eq.~\ref{eq:uncontrol}:
  
\begin{eqnarray}
  \frac{\Delta E}{E} = \delta k \cdot \left[ 1 - \frac{E}{\mathcal{E}(E)} \right]
\end{eqnarray}
  where $\mathcal{E}(E)\sim 100$ GeV in dilepton events. 
  The energy dependence in the classic calibration with single parameter in 
  Eq.~\ref{eq:runningunc} is reduced from $b/\mathcal{E}(E)$ to $\delta k$.
  The relative uncertainty as a function 
  of $E$ is shown in Fig.~\ref{fig:unc}.

\begin{figure}[h]
\begin{center}
\includegraphics[width=0.4\textwidth]{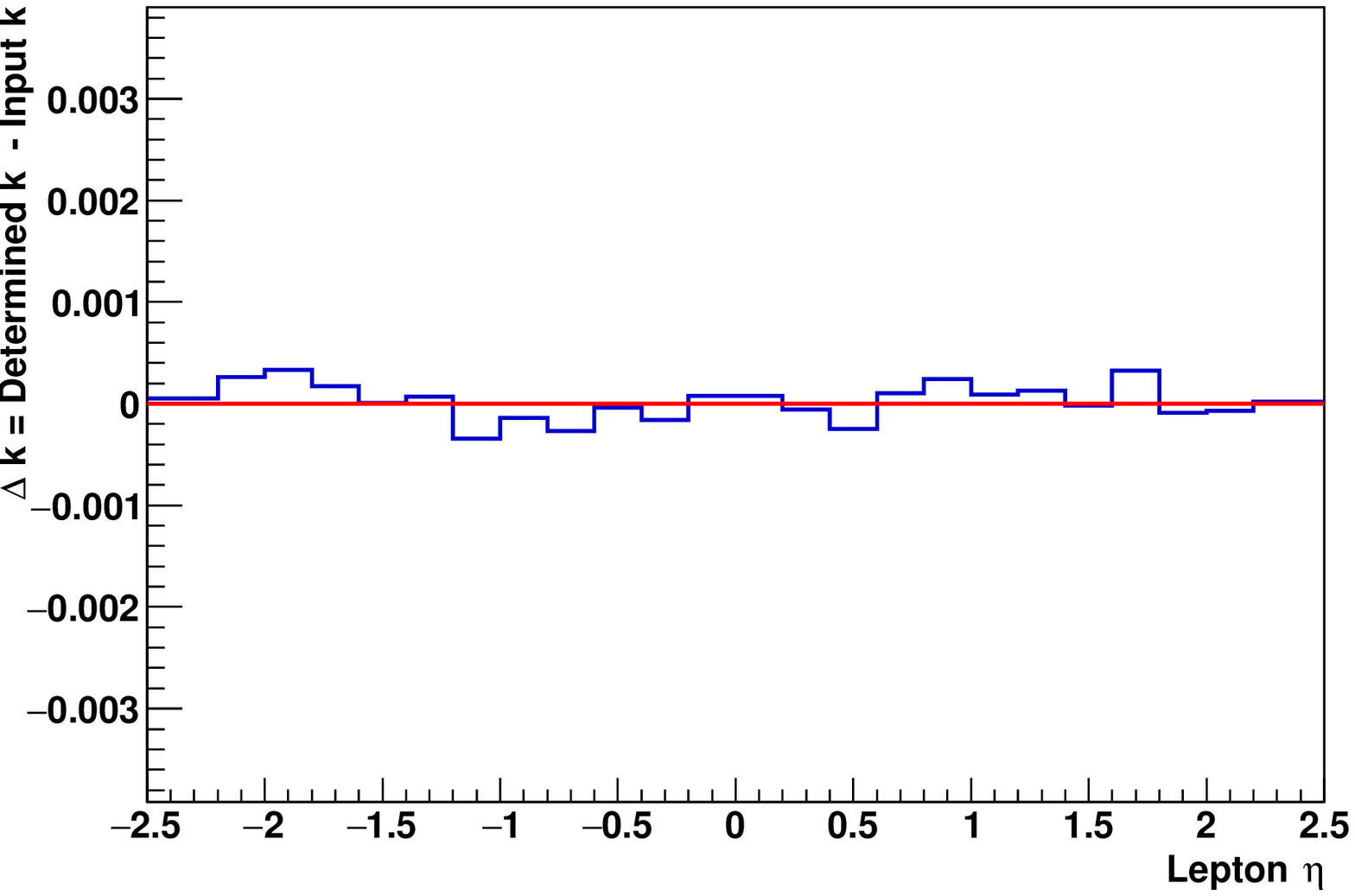}
\includegraphics[width=0.4\textwidth]{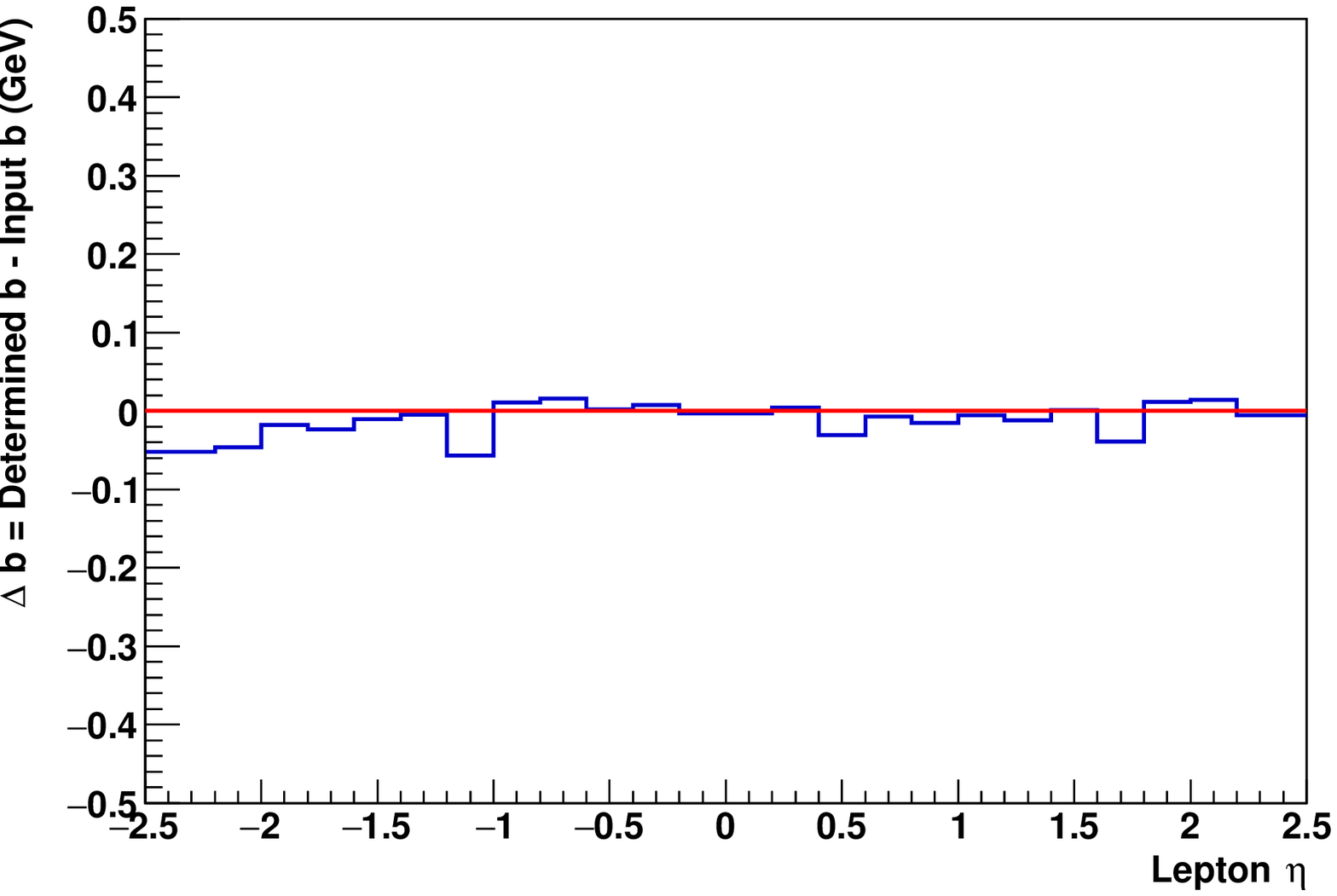}
\caption{\small $\delta k$ and $\delta b$ between the input values and the fitted values in large sample 
test. Relative uncertainty is 
smaller than 0.0005. $\delta b$ is in measurement of GeV.}
\label{fig:stattest}
\end{center}
\end{figure}

  Note that in the new calibration method, the dependence $\delta k$ can be further reduced with 
  larger data sample, while in the single scaling calibration, the running speed $b/\mathcal{E}(E)$ cannot 
  be reduced. To prove the uncertainty is dominated by statistical fluctuations, a generator level test using 
  larger samples with 1000M events has been done. $\delta k$ and $\delta b$ from the large sample test 
  are shown in Fig.~\ref{fig:stattest}. The relative uncertainty is smaller than 0.0005. The improvement 
  from 0.002 to 0.0005 is 
  consistent with the increase of statistics from 72M to 1000M ($\sqrt{1000/72}$).
  
\begin{figure}[h]
\begin{center}
\includegraphics[width=0.4\textwidth]{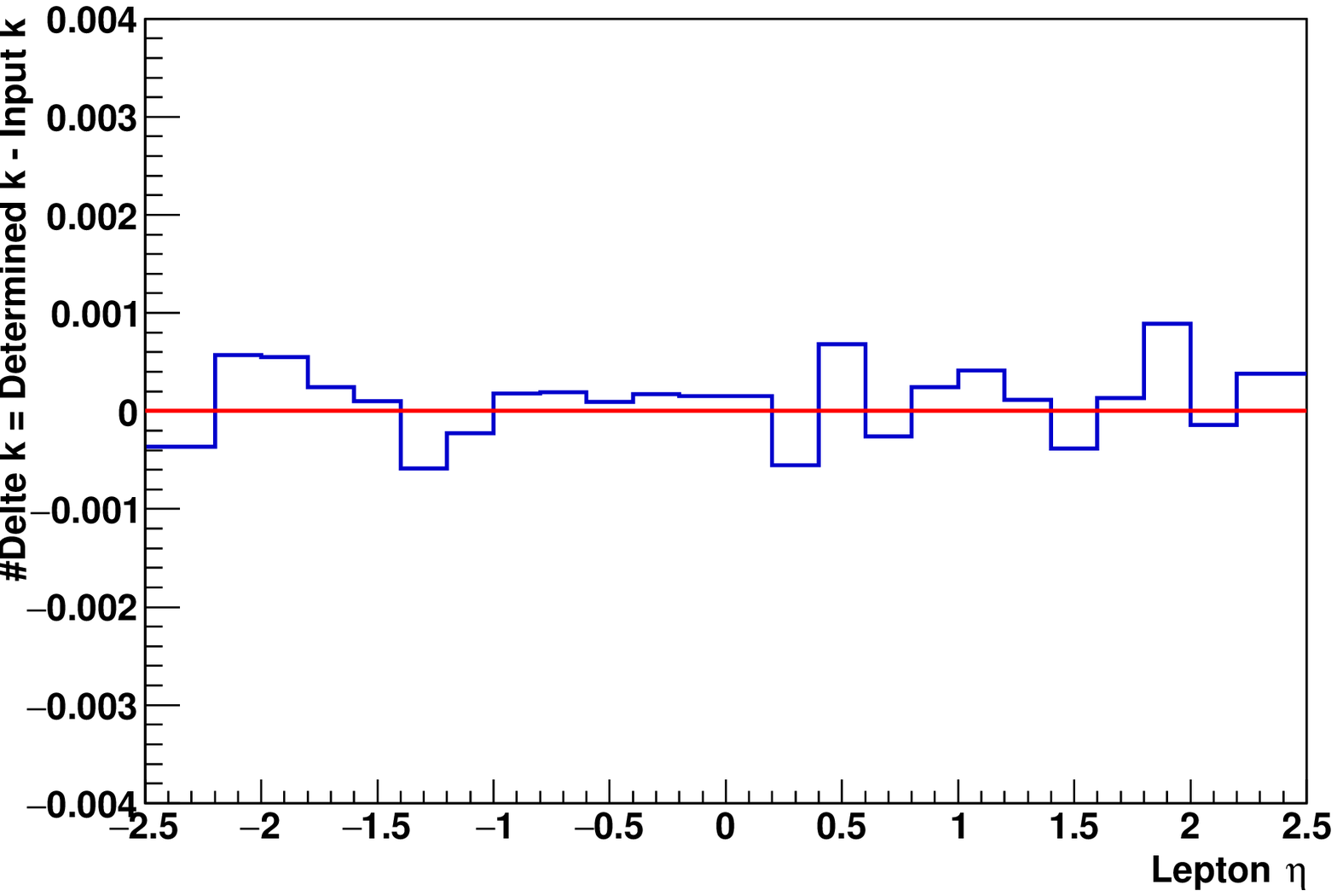}
\includegraphics[width=0.4\textwidth]{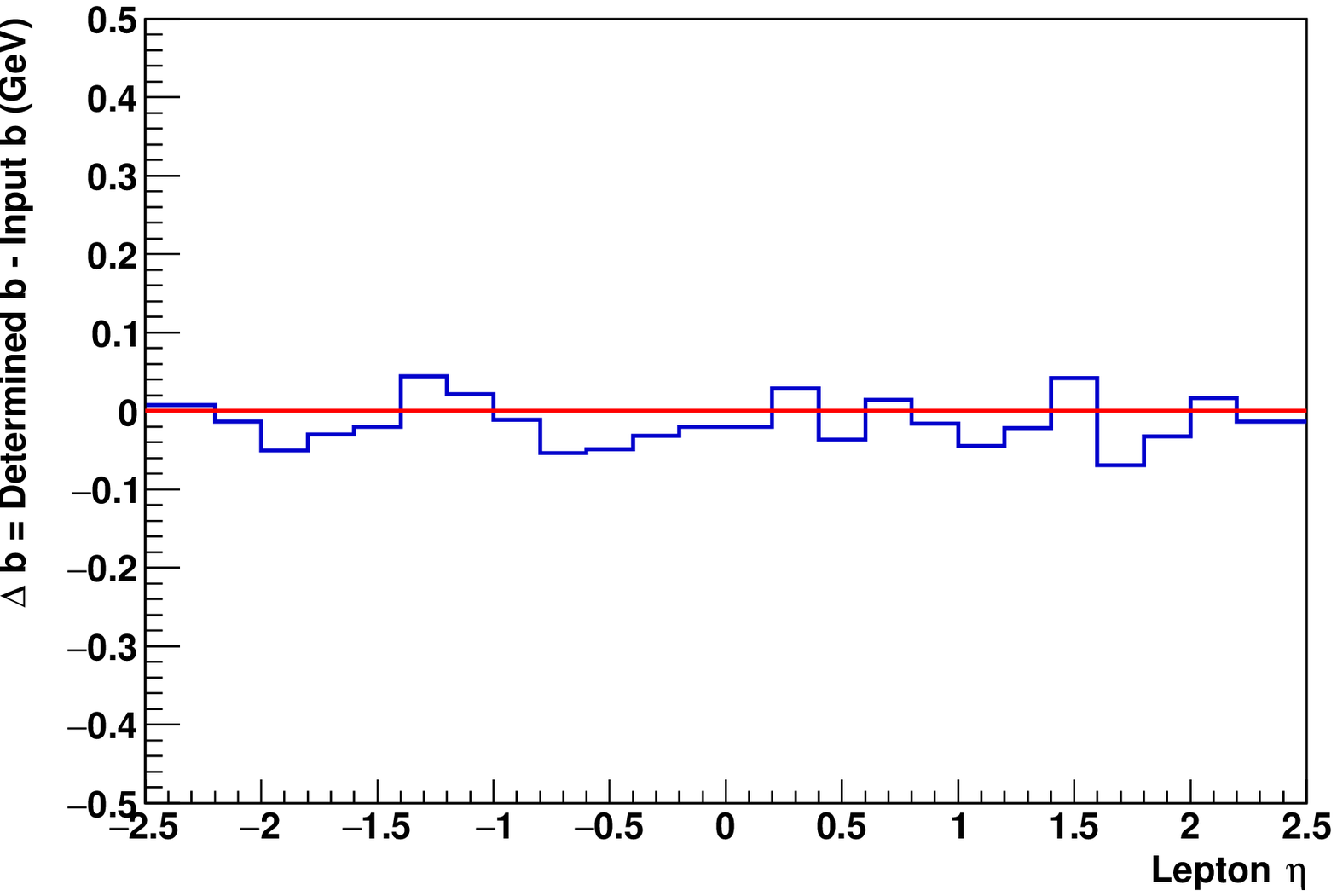}
\caption{\small $\delta k$ and $\delta b$ between the input values and the fitted values using PDF-differed samples. relative uncertainty is 
smaller than 0.0006. $\delta b$ is in measurement of GeV.}
\label{fig:PDFtest}
\end{center}
\end{figure}

  The calibration should be independent of PDFs. Another test using 1000 M large sample has been done to estimate the effect from 
  PDFs. In this test, the energy shifted sample is generated with NNPDF3.1 NLO No. 397, while the nominal sample 
  is generated with NNPDF3.1 LO No. 36. 
  Fig.~\ref{fig:PDFtest} shows the $k$ and $b$ values determined using the PDF-differed samples. 
  Even though these two PDF sets cause large energy difference as listed in Tab.~\ref{tab:pdftest}, $\delta k$ and $\delta b$ 
  are still less than $0.0006$, which is consistent with uncertainties in Fig.~\ref{fig:stattest}
  This test indicates the systematic uncertainty from PDFs is negligible compared to the statistical uncertainty. 

\section{IV. Alternative Methods}\label{sec:alternative}
\subsection{IV-A. Forward lepton calibration}
  The procedure described in section II and section III uses dilepton events with both leptons in the same $\eta$ region. However 
  at hadron collider experiments, dilepton events with both leptons in forward region with high $\eta$ are difficult to 
  be reconstructed. The coverage of the inner detector, the quality and efficiency of lepton measurement 
  are also limited due to the large contribution of backgrounds. To perform a forward lepton calibration, the calibration method is 
  modified so that we do not need events with both leptons in forward region.

  Assume the central region ($C$) where lepton $\eta$ is relatively low has already been calibrated using 
  the calibration method in section III. Separate it into multiple 
  subregions, denoted as $C_i$. For a given forward region ($F$) with high $\eta$, 
  the mass of dilepton events $C_i F$, with one lepton in $F$ and the other in $C_i$, 
  is written as:
  
\begin{eqnarray}
  E^F_\text{corr} &=& b_F + k_F \cdot E^F_\text{obs} \nonumber \\
  M^2_\text{true}[C_iF] &=& M^2_\text{corr}[C_iF] \nonumber \\
     &=& 2E^{C_i}E^{F}_\text{obs} (1-\cos\theta_{iF}) \nonumber \\
    &=& 2E^{C_i} (b_F + k_F\cdot E^{F}_\text{obs}) (1-\cos\theta_{iF})
\end{eqnarray}
  where $E^{C_i}$ is considered to be well calibrated. Then we can have a linear 
  relationship between $b_F$ and $k_F$:

\begin{eqnarray}
  b_F &=& \mathcal{A} \cdot k_F + \mathcal{B} \nonumber \\
     &=& \mathcal{E}(-E^{F}_\text{obs}) \cdot k_F + \mathcal{E}\left[\frac{M^2_\text{true}[C_iF]}{M^2_\text{obs}[C_iF]}\cdot E^F_\text{obs} \right]
\end{eqnarray}
  where $\mathcal{A}$ and $\mathcal{B}$ are determined by the mean of observed $F$ lepton 
  energy $E^F_\text{obs}$, the observed mass $M_\text{obs}[C_iF]$ and its true value 
  $M_\text{true}[C_iF]$. 
  Note that a mass constraint only provides $\mathcal{E}(M_\text{true}[C_iF])$, not 
  $M_\text{true}[C_iF]$ for each individual event. So  $\mathcal{B}$ is affected by covariance 
  between $M_\text{obs}[C_iF]$, $M_\text{true}[C_iF]$ and $E^F_\text{obs}$. 
  It can be observed in an alternative way.
  Fix $k_F$ with a 
  given value, then fit for the value of $b_F$ by minimizing the $\chi^2$ defined as:
  
\begin{eqnarray}
  \chi^2 = \frac{[ \mathcal{E}(M_\text{corr}[C_iF]) - \mathcal{E}(M_\text{true}[C_iF]) ]^2}{\sigma^2_M[C_iF]}.
\end{eqnarray}
  Note that with only one factor $b_F$ in the fitting, one can always achieve a good fitting result with low 
  $\chi^2$. 
  Then, 
  fix $k_F$ to other values and repeat the fitting for $b_F$. Finally we have a group 
  of $b_F$-$k_F$ pairs where the values of $k_F$ are given while the values of $b_F$ are 
  fitted. These pairs can be used to further fit the linear relationship between $b_F$ and $k_F$.
  
  For each $C_i$, the relationship between $b_F$ and $k_F$ can be observed independently, with different 
  slope $\mathcal{A}$ and offset $\mathcal{B}$. These 
  $b_F$-$k_F$ lines should have one intersection point, corresponding to the determined value of 
  $b_F$ and $k_F$.
  
\begin{figure}[h]
\begin{center}
\includegraphics[width=0.45\textwidth]{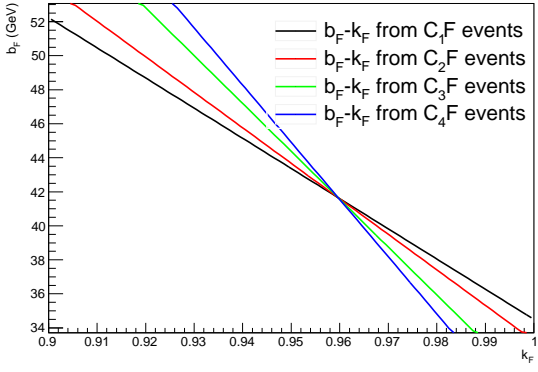}
\caption{\small Observed $b_F$-$k_F$ lines in each $C_iF$ categories. The intersection point corresponds 
to the determined values of $b_F$ and $k_F$.}
\label{fig:forwardlepton}
\end{center}
\end{figure}

  A closure test is performed as an example using the 72M sample. The $F$ region is given as $2.5<\eta<3.0$. 
  Parameters $k_F = 0.960$ and $b_F = -41.651$ GeV, are applied to the $F$ lepton energy. 
  The $C$ region is 
  separated into four subsamples, given in Tab.~\ref{tab:ForwardSubsample}. For each $C_i$, 
  the observed slope $\mathcal{A}$ and $\mathcal{B}$ are also listed. Fig.~\ref{fig:forwardlepton} 
  shows the four observed $b_F$-$k_F$ lines. Their intersection point corresponds to the determined values of 
  $k_F = 0.966$ and $b_F = -40.116$ GeV. Compared to the injected $k_F$ and $b_F$ value, the running factor from  
  Eq.~\ref{eq:uncontrol} is $\delta k_F = 0.006$. It is much smaller than the running factor 
  of $b_F / \mathcal{E}(E^F_\text{obs}) \sim 40 / 200 = 0.25$ in Eq.~\ref{eq:runningunc} 
  for a single factor calibration.
  
\begin{table}[h]
\begin{center}
\begin{tabular}{l|c|c|c}
\hline \hline
       &  $\eta$ range & $\mathcal{A}$ from $C_iF$(GeV) & $\mathcal{B}$ from $C_iF$ (GeV) \\
\hline
 $C_1$  & [-2.5, 0] & -177.327 & 211.830 \\
\hline
 $C_2$  & [0, 0.8]  & -208.678 & 241.918 \\
\hline
 $C_3$  & [0.8, 2.0]  & -281.295 & 311.607\\
\hline
 $C_4$  & [2.0, 2.5]  & -336.097 & 364.201 \\
\hline
\hline \hline
\end{tabular}
\caption{\small An example of forward lepton calibration.}
\label{tab:ForwardSubsample}
\end{center}
\end{table}

  In principle, $k_F$ and $b_F$ can be determined with two $C_i$ regions. But two lines always have one 
  intersection point, even if there is bias in the $b_F$-$k_F$ relationship observation. It is good to have more 
  than two $C_i$ regions, so that the calibration can be tested by checking whether these $b_F$-$k_F$ lines 
  intersect at one point.

\subsection{IV-B. Muon calibration}

  Muon momentum is measured from the fitted curvature of muon tracks, which has dependence with 
  muon charge due to detector misalignment~\cite{misalignment}. 
  When charge dependence is introduced, we have 12 parameters of $k^\pm$ and $b^\pm$ in a group 
  of $H$, $M$ and $L$ regions, but only 9 mass constraints from subsamples $H^+H^-$, $H^+M^-$, $H^+L^-$, 
  $M^+H^-$, $M^+M^-$, $M^+L^-$, $L^+H^-$, $L^+M^-$ and $L^+L^-$. We introduce 3 additional constraints 
  by observing variable $R$ defined as:
  
 \begin{eqnarray}
  R_L &=& \frac{\mathcal{E}(E^{L^+}_{L^+L^-})}{\mathcal{E}(E^{L^-}_{L^+L^-})} \nonumber \\
  R_M &=& \frac{\mathcal{E}(E^{M^+}_{M^+M^-})}{\mathcal{E}(E^{M^-}_{M^+M^-})} \nonumber \\  
  R_H &=& \frac{\mathcal{E}(E^{H^+}_{H^+H^-})}{\mathcal{E}(E^{H^-}_{H^+H^-})}
 \end{eqnarray}
  where $E^{\alpha^\pm}_{\alpha^+\alpha^-}$ $(\alpha=L, M, H)$ are energy of $\mu^\pm$ in events $\alpha^+\alpha^-$.
  The constraints can be expressed as requiring $R^\text{corr}_\alpha$ after correction 
  to be consistent with $R^\text{true}_\alpha$ calculated from true energy mean $\mathcal{E}(E_\text{true})$:
  
\begin{eqnarray}
  R^\text{obs}_\alpha \rightarrow R^\text{corr}_\alpha = R^\text{true}_\alpha.
\end{eqnarray} 
  Note that we use the ratio of energy instead of the energy itself to avoid effects from PDFs and QCD calculation.  
  Such effects are less significant in the ratio since energies of $\mu^+$ and $\mu^-$ are correlated. 
  
  A procedure of reducing correlations between $k^\pm$ and $b^\pm$ is still needed. The procedure 
  is similar to the electron case with slight changes. 
  The dimuon masses are:
  
\begin{eqnarray}
  M^2_\text{true} &=& 2(k^+ E^+_\text{obs} + b^+)(k^- E^-_\text{obs} + b^-) (1-\cos\theta_{+-}) \nonumber \\
  M^2_\text{obs} &=& 2E^+_\text{obs} E^-_\text{obs} (1-\cos\theta_{+-}).
\end{eqnarray}
  Thus we have
  
\begin{eqnarray}
  \frac{M_\text{true}}{M_\text{obs}} = \sqrt{ \frac{k^+ E^+_\text{obs} + b^+}{E^+_\text{obs}}\cdot \frac{k^- E^-_\text{obs} + b^-}{E^-_\text{obs}} }.
\end{eqnarray}
  As shown in section II-C, a correction $\epsilon$ should be added due 
  to the covariance between $M$, $E^+$ and $E^-$ when calculating for mean values:

\begin{footnotesize}  
\begin{eqnarray}
  \sqrt{ \frac{k^+ \mathcal{E}(E^+_\text{obs}) + b^+}{\mathcal{E}(E^+_\text{obs})}\cdot \frac{k^- \mathcal{E}(E^-_\text{obs}) + b^-}{\mathcal{E}(E^-_\text{obs})} } = \frac{\mathcal{E}(M_\text{true})}{\mathcal{E}(M_\text{obs})} + \epsilon.
\end{eqnarray}
\end{footnotesize}
  Defining variables of $R$ as 
  
\begin{eqnarray}
  R^\text{obs} &=& \frac{\mathcal{E}(E^+_\text{obs})}{\mathcal{E}(E^-_\text{obs})} \nonumber \\
  R^\text{true} &=& \frac{k^+\mathcal{E}(E^+_\text{obs}) + b^+}{k^-\mathcal{E}(E^-_\text{obs})+b^-}
\end{eqnarray}
  we have:

\begin{footnotesize}
\begin{eqnarray}\label{eq:muoncorrelation}
  k^+ \mathcal{E}(E^+_\text{obs}) + b^+ &=& \mathcal{E}(E^+_\text{obs})\cdot \sqrt{\frac{R^\text{true}}{R^\text{obs}} }\cdot \left[ \frac{\mathcal{E}(M_\text{true})}{\mathcal{E}(M_\text{obs})} + \epsilon \right] \nonumber \\
  k^- \mathcal{E}(E^-_\text{obs}) + b^- &=& \mathcal{E}(E^-_\text{obs})  \cdot \sqrt{\frac{R^\text{true}}{R^\text{obs}}}\cdot \left[ \frac{\mathcal{E}(M_\text{true})}{\mathcal{E}(M_\text{obs})} + \epsilon \right].
\end{eqnarray}
\end{footnotesize}  

  Eq.~\ref{eq:muoncorrelation} allows us to use $k^\pm$ and $\epsilon$ to represent $b^\pm$. These relationships can be observed using $\alpha^+\alpha^-$ events ($\alpha=L, M, H$):
  
\begin{eqnarray}
  k^\pm_\alpha \mathcal{E}(E^{\alpha^\pm}_{\alpha^+\alpha^-}[\text{obs}]) &+& b^\pm_\alpha = \mathcal{E}(E^{\alpha^\pm}_{\alpha^+\alpha^-}[\text{obs}])\cdot \sqrt{\frac{R^\text{true}_\alpha}{R^\text{obs}_\alpha} } \nonumber \\
  & & \times \left[ \frac{\mathcal{E}(M_\text{true}[\alpha^+\alpha^-])}{\mathcal{E}(M_\text{obs}[\alpha^+\alpha^-])} + \epsilon_\alpha \right].
\end{eqnarray}
  $k^\pm_\alpha$ and $\epsilon_\alpha$ can be further determined by 
  minimizing the $\chi^2$ defined as 

\begin{footnotesize}  
\begin{eqnarray}
  \chi^2 &=& \sum_{\alpha^+\beta^-}\frac{\left[ \mathcal{E}(M_\text{corr}[\alpha^+\beta^-]) - \mathcal{E}(M_\text{true}[\alpha^+\beta^-]) \right]^2}{\sigma^2_M[\alpha^+\beta^-] } \nonumber \\
     &+&  \sum_{\alpha^-\beta^+}\frac{\left[ \mathcal{E}(M_\text{corr}[\alpha^-\beta^+]) - \mathcal{E}(M_\text{true}[\alpha^-\beta^+]) \right]^2}{\sigma^2_M[\alpha^-\beta^+] } \nonumber \\
     &+& \sum_\alpha \frac{\left[ R^\text{obs}_\alpha - R^\text{true}_\alpha \right]^2}{\sigma^2_R[\alpha]}
\end{eqnarray}
\end{footnotesize}
  where $\alpha, \beta=L,M,H$ refer to nine mass constraints and three energy ratio constraints. $\sigma_M$ is the uncertainty on observed mass mean, and $\sigma_R$ is the uncertainty 
  on $R$. 
  
  This method is tested using the 72 M samples. 
  As an example, Twelve parameterrs of $k^\pm$ and $b^\pm$ for three $\eta$ regions, [-2.5, -2.2], [-1.6, -1.4] and [0, 0.2], are applied 
  to shift the muon energy. Their values are randomly given. The muon calibration is applied to determine values of the factors using 
  a nominal sample. 
  The input values and determined values are listed in Tab.~\ref{tab:muoncalib}.  The average uncertainty of $\delta k \sim 0.005$ is larger than the uncertainty 
  in Fig.~\ref{fig:deltaPara}, because events are separated into smaller subsamples with larger statistical fluctuations. Another reason is the uncertainty 
  $R$ variable observation using muon energy has larger uncertainties.
  
\begin{table}[h]
\begin{center}
\begin{tabular}{l|c|c|c|c|}
\hline \hline
       & Input & Determined & Input & Determined \\
       & $k$ & $k$  & $b$ (GeV) & $b$ (GeV) \\
\hline
[0, 0.2] & & & & \\
$\mu^-$ &  1.0258 & 1.0199 & 2.341 & 2.609 \\
$\mu^+$ & 1.0133 & 1.0157 & 2.841 & 2.729 \\
\hline
[-1.6, -1.4] & & & & \\
$\mu^-$ &  0.9889 & 0.9876 & -2.216 & -2.005 \\
$\mu^+$ & 0.9736 & 0.9783 & -1.916 & -2.4569 \\
\hline
[-2.5, -2.2] & & & & \\
$\mu^-$ &  0.9810 & 0.9783 & -2.761 & -1.969\\
$\mu^+$ & 0.9706 & 0.9775 & -1.561 & -3.276 \\
\hline \hline
\end{tabular}
\caption{\small Input values and determined values of $k^\pm$ and $b^\pm$ in $\eta$ regions of one group.}
\label{tab:muoncalib}
\end{center}
\end{table}
  
\section{V. Further Discussion}\label{sec:further}
\subsection{V-A. $\eta$-dependence in mass observation}

\begin{figure}[!hbt]
\begin{center}
\includegraphics[width=0.45\textwidth]{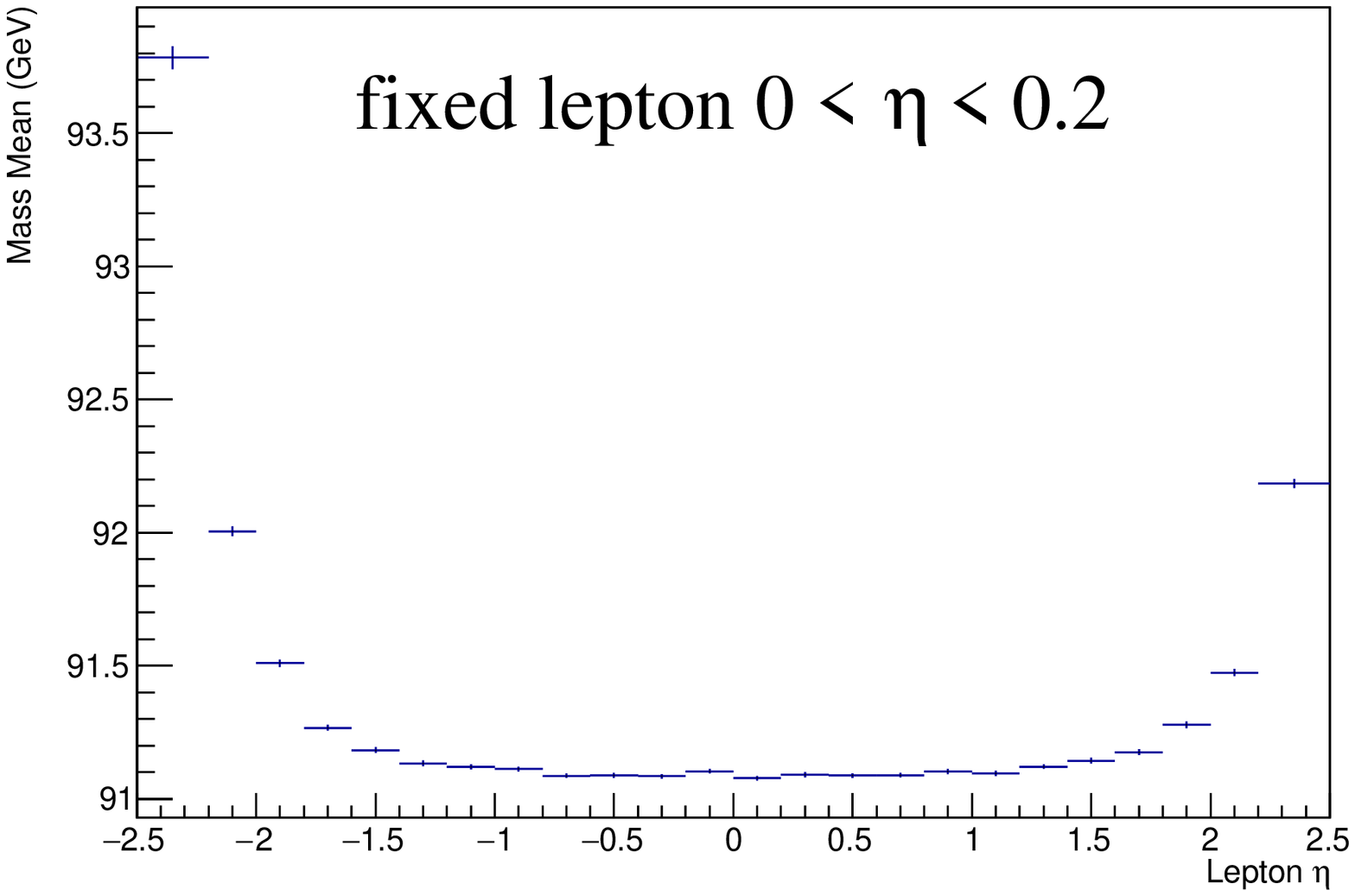}
\includegraphics[width=0.45\textwidth]{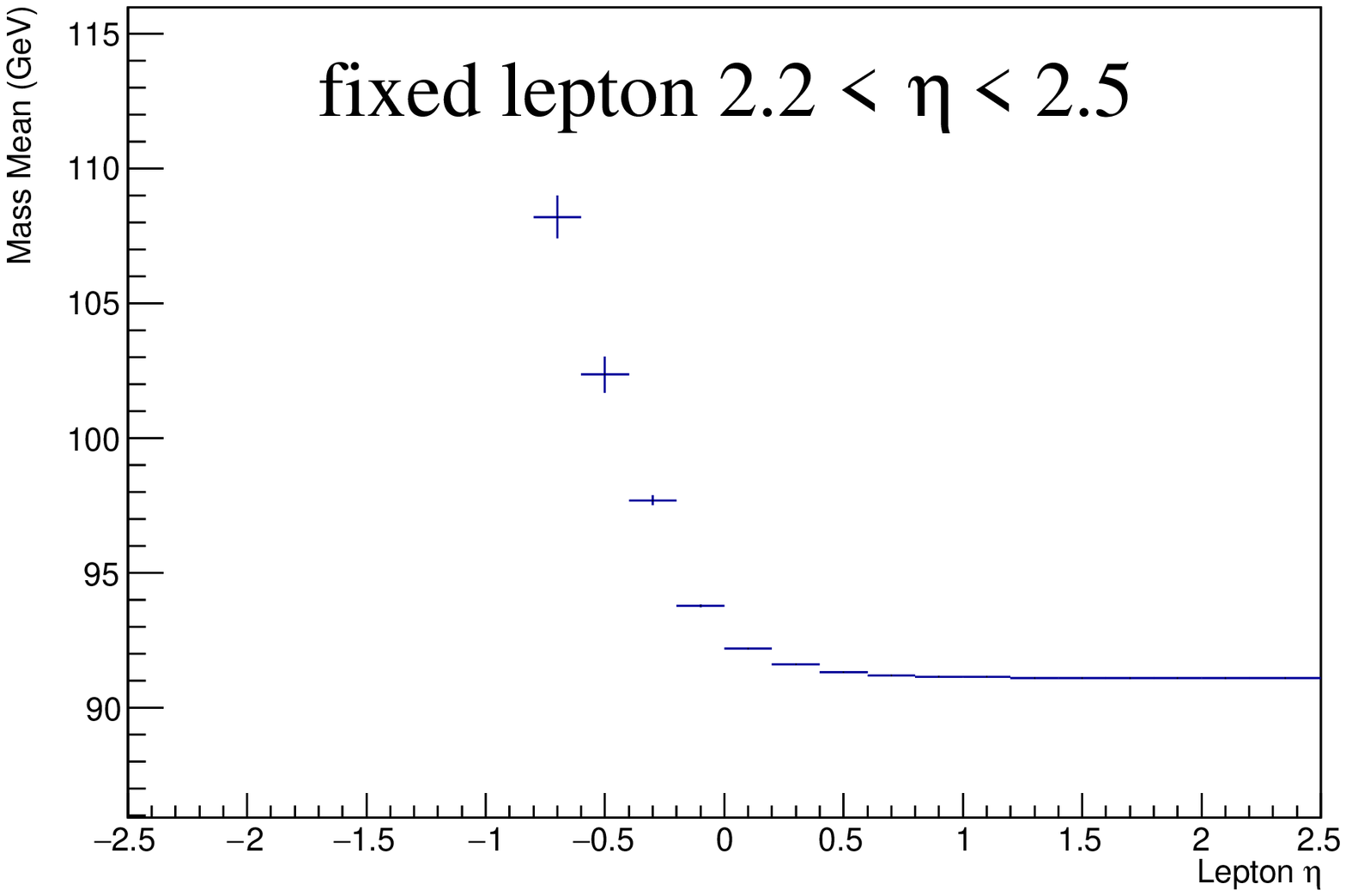}
\caption{\small Energy of the lepton in $Z\rightarrow \ell^+\ell^-$ events. The Y-axis is the mean of the dilepton mass spectrum 
with one lepton fixed in $\eta$ region [0, 0.2] (top) 
and [2.2, 2.5] (bottom). The X-axis is the $\eta$ 
of the other lepton in the dilepton events. Note that some bins are empty because there is no event if $\Delta \eta$ between 
two leptons is too large.}
\label{fig:zmass}
\end{center}
\end{figure}

  The mean value of mass spectra is a constant only in full phase space of final state leptons 
  from $Z$ boson decay. In the calibration, dilepton events are applied with lepton $p_T$ and $\eta$ cuts. 
  The mean value of dilepton mass spectrum in each subsample corresponds to a cut-phase space, 
  thus has dependence of lepton $\eta$. 
  When the opening angle between the two leptons in 
  the $Z/\gamma^* \rightarrow \ell^+\ell^-$ events is large, at least one of the lepton $p_T$ is lower. Therefore a lepton $p_T$ cut 
  tend to remove low mass events and keep high mass events. As a result, the mean of the mass is shifted 
  by the cut. Such effect can be very large, as shown in Fig.~\ref{fig:zmass}. 
  
  To consider such effect, when performing an absolute calibration, $\mathcal{E}(M_\text{true})$ must be acquired from 
  a generator-level sample with lepton $p_T$ and $\eta$ cut same as that applied in data and MC simulations.

\subsection{V-B. Correlation with resolution}

\begin{figure}[!hbt]
\begin{center}
\includegraphics[width=0.45\textwidth]{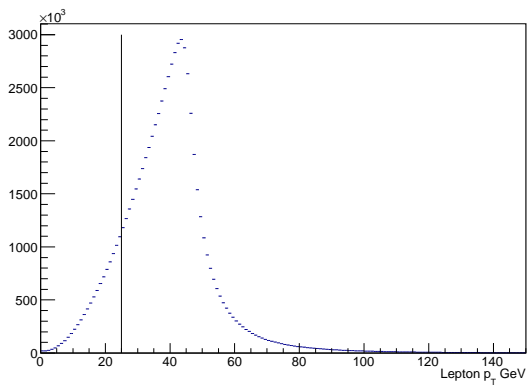}
\caption{\small Distribution of lepton $p_T$ from $Z/\gamma^*\rightarrow \ell^+\ell^-$ events before any $p_T$ cut.}
\label{fig:leptonpT}
\end{center}
\end{figure}

  The calibration on lepton energy scale has correlation with energy resolution, which is also from lepton $p_T$ cut. The 
  lepton $p_T$ spectra from $Z/\gamma^* \rightarrow \ell^+\ell^-$ events is shown in Fig.~\ref{fig:leptonpT}. Due to a 
  peak structure, events with lepton $p_T$ larger than the cut value are more than events with lepton $p_T$ lower than 
  than cut value. Even if the resolution itself is symmetrical, it smears more events from high $p_T$ to low $p_T$. 
  As a result, a cut applied on the reconstructed  lepton $p_T$ is slightly rejecting more high $p_T$ leptons. 
  It will cause a non linear relationship between $E_\text{obs}$ and $E_\text{true}$ in the selected sample. 
  It is purely a statistical effect, as if the $p_T$ cut is selecting data sample with bias.  

\begin{figure}[!hbt]
\begin{center}
\includegraphics[width=0.45\textwidth]{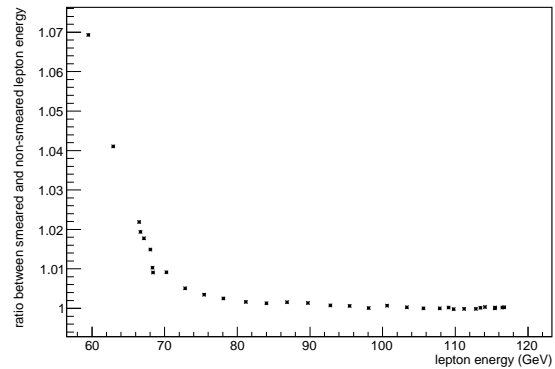}
\caption{\small Ratio of smeared lepton energy and corresponding true energy as a function of lepton energy.}
\label{fig:nonlinear}
\end{center}
\end{figure}

  Fig.~\ref{fig:nonlinear} shows an example of the resolution effect. A $5\%$ gaussian smearing ($N(0, 0.05^2)$) is applied to the 
  lepton energy in the $Z/\gamma^* \rightarrow \ell^+\ell^-$ sample. A $p_T>25$ GeV cut is then applied to the lepton 
  $p_T$ after smearing. As we can see, the ratio of average lepton energy before and after smearing is not a constant, even 
  if the smearing itself does not change the average of lepton energy. If the resolution is not symmetrical, such effect will be 
  even larger. This effect has to be considered in the energy scale calibration. $\mathcal{E}(E_\text{true})$ and $\mathcal{E}(E_\text{true})$ 
  must be observed with consistent resolution smearing, which requires a good modeling of the detector resolution. 
  
\section{VI. Summary}\label{sec:summary}

  In conclusion, we described a new calibration method using $Z\rightarrow \ell^+\ell^-$ method. The new method allows 
  offset terms in the calibration function and can precisely determine the values of the 
  parameters. The method first introduces multiple-mass constraints by separating the $Z\rightarrow \ell^+\ell^-$ events 
  according to the opening angle between leptons. Then, a step by step fitting procedure is used to 
  reduce the remaining correlation between parameters. 
  A generator level test shows that the precision of the $k$ and $b$ parameters determined by this method is 
  around $0.2\%$, and the precision of the energy calibration is $<10^{-4}$, with a data sample equivalent to 
  35 fb$^{-1}$ data collected by the ATLAS or the CMS detector at the LHC 13 TeV. The uncertainty is dominated by the 
  data sample which can be further reduced with more events. With slight modification, the calibration can be 
  used for forward lepton calibration where dilepton events must have at least one lepton in central region, and 
  for muon calibration where charge-dependence must be introduced.
  This method uses only information of the reconstructed dilepton mass. The precision is much higher than the 
  classic single-parameter calibration. It is much faster and easier than performing a perfect detector 
  simulation, or an advanced but time-consuming fitting technique. 

\section{Acknowledgements}
\begin{acknowledgements}
  We thank Professor Paul. Grannis for his help in providing comments and general 
  suggestions for the summarization of the new method. 
\end{acknowledgements}

\end{document}